\documentclass[11pt,letterpaper]{article}
\usepackage{hyperref}
\usepackage{latexsym}
\usepackage{amssymb,amsfonts,amsmath}
\usepackage{graphicx}
\usepackage{indentfirst}
\usepackage{amsmath}
\usepackage{amsfonts}
\usepackage{amssymb}%
\usepackage{comment}
\usepackage{mathtools}
\usepackage{amsthm}
\usepackage{enumerate}
\usepackage{extarrows}
\usepackage{relsize}

\usepackage[cmtip,all]{xy}
\usepackage{verbatim}
\usepackage{tikz-cd}
\usepackage{mathtools}
\usepackage{mathrsfs}
\usepackage{geometry}

\DeclareMathAlphabet{\mathpzc}{OT1}{pzc}{m}{it}

\newcommand{\beq}{\begin{equation}}
\newcommand{\eeq}{\end{equation}}
\newcommand{\bear}{\begin{eqnarray}}
\newcommand{\eear}{\end{eqnarray}}

\newcommand*{\defeq}{\mathrel{\vcenter{\baselineskip0.5ex \lineskiplimit0pt
                     \hbox{\scriptsize.}\hbox{\scriptsize.}}}%
                     =}

\ifx\pdfoutput\undefined
\else
\fi
\hypersetup{colorlinks=false,bookmarksopen,bookmarksnumbered,citecolor=blue,
pdfstartview=FitH}

\parskip=\medskipamount

\def\a{\alpha}

\def\b{\beta}

\newcommand{\be}{\begin{equation}}
\newcommand{\ee}{\end{equation}}
\newcommand{\bea}{\begin{eqnarray}}
\newcommand{\eea}{\end{eqnarray}}

\newcommand{\ba}{\begin{array}}
\newcommand{\ea}{\end{array}}

\def\double #1{#1{\hbox{\kern-2pt $#1$}}}

\newcommand{\bsubeq}{\begin{subequations}}
\newcommand{\esubeq}{\end{subequations}}

\newcommand{\virgolette}{``}

\theoremstyle{definition}

\theoremstyle{plain}

\theoremstyle{remark}

\begin{document}

\begin{titlepage}
\begin{flushright}
\par\end{flushright}
\vskip 0.5cm
\begin{center}
\textbf{\LARGE \bf Cohomology of Lie Superalgebras}\\
\vskip .2cm
\textbf{\Large \bf Forms, Pseudoforms, and Integral Forms}

\vskip .5cm

%\large {\bf R.~Catenacci}$^{~a,b,c,}$\footnote{roberto.catenacci@uniupo.it}
\large {\bf C.~A.~Cremonini}$^{~a,b,}$\footnote{carlo.alberto.cremonini@gmail.com} and 
\large {\bf P.~A.~Grassi}$^{~b, c,d,}$\footnote{pietro.grassi@uniupo.it},
%\large {\bf S.~Noja}$^{~g,}$\footnote{noja@mathi.uni-heidelberg.de}\,.  

\vskip .5cm {
\small
%\centerline{$^{(a)}${\it Gruppo Nazionale di Fisica Matematica, InDAM, Piazzale Aldo Moro 5, 00185, Roma}}
\centerline{$^{(a)}$ \it Galileo Galilei Institute for Theoretical Physics,}
\centerline{\it Largo Fermi 2, 50125, Firenze, Italy}
\centerline{$^{(b)}$ \it INFN, Sezione di Torino}
\centerline{\it via P.~Giuria 1, 10125 Torino, Italy} 
\centerline{$^{(c)}$
\it Dipartimento di Scienze e Innovazione Tecnologica (DiSIT),} 
\centerline{\it Universit\`a del Piemonte Orientale,} 
\centerline{\it viale T.~Michel, 11, 15121 Alessandria, Italy}
\centerline{$^{(d)}$
\it Arnold-Regge Center,}
\centerline{\it 
via P.~Giuria 1,  10125 Torino, Italy}
}
\end{center}

\begin{abstract}

We study the cohomology of Lie superalgebras for the full complex of forms: superforms, pseudoforms and integral forms. We use the technique of spectral sequences to abstractly compute the Chevalley-Eilenberg cohomology. We first focus on the superalgebra $\mathfrak{osp}(2|2)$ and show that there exist non-empty cohomology spaces among pseudoforms related to sub-superalgebras. We then extend some classical theorems by Koszul, as to include pseudoforms and integral forms. Further, we conjecture that the algebraic Poincar\'e duality extends to Lie superalgebras, as long as all the complexes of forms are taken into account and we prove that this holds true for $\mathfrak{osp}(2|2)$. We finally construct the cohomology representatives explicitly by using a distributional realisation of pseudoforms and integral forms. On one hand, these results show that the cohomology of Lie superalgebras is actually larger than expected, whereas one restricts to superforms only; on the other hand, we show the emergence of completely new cohomology classes represented by pseudoforms. These classes realise as integral form classes of sub-superstructures.

\end{abstract}

\vfill{}
\vspace{1.5cm}
\end{titlepage}

%\maketitle
\setcounter{footnote}{0}
\tableofcontents

\vfill
\eject
\section{Introduction} \setcounter{equation}{0}

The advent of Lie superalgebras \cite{Kac:1977em} was useful not only for new mathematical studies, but also 
in the physical realm of supersymmetric theories. Since then, several applications and discoveries have used the natural 
generalization of the Lie algebra framework to graded Lie algebras, and in particular several studies were carried out to classify all possible interesting examples 
of Lie superalgebras \cite{Kac:1977em,Manin:1988ds,Leites:1982ia}. As for Lie algebras, one of these classification techniques involves 
algebraic invariants and the correspondent Chevalley-Eilenberg cohomology \cite{CE,Koszul}. Therefore, it 
appeared straightforward to translate the Chevalley-Eilenberg analysis to the context of Lie superalgebras, as in the pioneering work \cite{Fuks}. 

The main difference between Lie superalgebras and Lie algebras is presence of anti-commuting (odd) generators  which drastically enrich the results.
For a complete dictionary on superalgebras we refer to \cite{Frappat}. 
Associated to those anti-commuting generators, there are the dual Maurer-Cartan forms which appear as commuting differential forms, in contrast 
to the usual anti-commuting 1-forms associated to the even generators. Using the Maurer-Cartan equations,  the Chevalley-Eilenberg cohomology is easily defined by following \cite{CE,Koszul} on the 
space of differential forms, namely on \emph{superforms}. General theorems for Lie algebras (see, e.g., \cite{GHV}) point out that there is 1-1 correspondence between the algebraic invariants and the  cohomology classes (in the present work, we will deal only with Casimir invariants and CE-cohomology with the trivial module $\mathbb{R}$). 
For Lie superalgebras, the crucial point 
is that the full cohomology does not coincide with the CE-cohomology of superforms only, but rather it embraces new 
type of forms which are known, in the super-geometric context, as {\it integral forms} and {\it pseudoforms} \cite{lei,CGN,Manin:1988ds,Witten}. The 
former have been introduced on supermanifolds to implement a meaningful integration theory \cite{Manin:1988ds,lei} 
and their formal properties were studied in the seminal papers \cite{BS,Belopolsky:1996cy,Belopolsky:1997bg,Belopolsky:1997jz,FMS,VV}, using an explicit distributional realisation. These new forms define a 
new complex on which the CE differential can be defined. This construction has been explored in \cite{CCGN2} where all the details are given. 

The superform cohomology has been well known since the '80s (see, e.g., \cite{Fuks}) and, recently, has been explored in several works (see, e.g., \cite{boe,Lebedev:2004mq,Scheunert:1997ks,pro,Su:2019znq}), but 
the exploration of the full cohomology has never been carried out: this is the aim of the present work.  We will use as a guiding example the superalgebra $\mathfrak{osp}(2|2)$, which is the smallest case where pseudoforms arise as cohomology classes. The analysis of this example is multi-purposed: first, we give the description complete cohomology for the whole complex of forms; second, we show how pseudoforms arise naturally, at least in this algebraic context.

This second result is far from trivial, since despite their recent use in physical contexts (and the formal properties they are supported with) lead to unexpected results (see, e.g., \cite{CGN,CGN2,CCGN,CG,CG2}), their full mathematical understanding is still lacking. In this paper we want to emphasise, at least in an algebraic context, how pseudoforms are strictly related to the existence of sub-superstructures. On a geometric setting, this is intuitively suggested (e.g., in \cite{Witten}) by introducing sub-supermanifolds with non-trivial odd codimension. In that paper, the author considers as a toy example $\mathbb{R}^{(0|*2)}$, locally parametrised by the two odd coordinates $\theta^1$ and $\theta^2$; he then uses the \virgolette odd constraint" $a \theta^1 + b \theta^2 = 0 , a,b \in \mathbb{C}$, to define a submanifold with odd codimension 1, and shows that the correct object to be integrated on such submanifold is actually a pseudoform. In that paper the author described also the very useful distributional realisation of pseudoforms and integral forms, which is particularly powerful for explicit calculations and physical applications (see also \cite{Wit2}). On the algebraic setting, the sub-superstructures are already built-in (at least in general): the sub-superalgebras. We will leverage on these structures to show how pseudoforms emerge in a natural way. In particular, we will show that sub-superalgebras play the analogous role to the constraints and that pseudoforms are related to the integral forms defined on sub-superalgebras and supercosets. The distributional realisation can be used, once again, as a powerful tool for inspiring calculations and suggesting results.

We will calculate the cohomology in two ways: in the first section we will consider the Koszul spectral sequence \cite{Koszul} (i.e., Hochschild-Serre spectral sequence \cite{HocSer} with trivial module) for $\mathfrak{osp}(2|2)$. In particular, we will infer pseudoforms as integral forms of the supercoset $\mathfrak{osp} \left( 2|2 \right) / \mathfrak{osp} \left( 1|2 \right)$ and of the sub-superalgebra $\mathfrak{osp} \left( 1|2 \right)$. In the second section we will introduce some general constructions, starting from the definition of a generalised spectral sequence for Lie superalgebras based on two filtrations. We will show that these two filtrations induce two inequivalent sectors of the first page of the spectral sequence when considering Lie sub-superalgebras, but reduce to a single sector when considering purely even Lie sub-algebras. We will also comment on some simplifications that occur for superform and integral form complexes. We will conclude the section with the proofs of general theorems that extend classical theorems for Lie algebra (see \cite{Koszul}) to the super setting, for any form  complex. In the third section we will show how pseudoforms of the main example $\mathfrak{osp}(2|2)$ emerge by using the explicit distributional realisation, when considering the purely odd coset $\mathfrak{osp}(2|2) / \mathfrak{sp} (2) \times \mathfrak{so}(2)$ and we will then construct invariant pseudoforms explicitly. We will support explicit calculations with Poincar\'e polynomials, whose use in this algebraic setting is briefly reviewed. In the two appendices we recall the basic rules for manipulating the distributional realisation of pseudoforms and integral forms and an easy example for calculations with Poincar\'e polynomials and cosets.

\section{Cohomology via Spectral Sequences}

In this section, we generalize the Koszul spectral sequence to compute pseudoform cohomology of the Lie superalgebra $\mathfrak{osp}(2|2)$ via the associated supercoset $\mathfrak{osp}(2|2) / \mathfrak{osp}(1|2)$.

By denoting with $Z,H,E^{\pm\pm}$ the bosonic generators and with $F^\pm, \bar{F}^\pm$ the fermionic ones, the non trivial (anti-)commutation relations of $\mathfrak{osp}(2|2)$ (see e.g., \cite{Frappat}\footnote{With respect to the fermionic generators of \cite{Frappat}, we used the combinations $\displaystyle \frac{1}{2} \left( F^\pm \pm \bar{F}^\pm \right) $ and $\displaystyle \frac{1}{2} \left( F^\pm \mp \bar{F}^\pm \right) $ in order to make the $\mathfrak{osp}(1|2)$ sub-superalgebra manifest.}) are
\begin{eqnarray}
	\label{VSPA} && \left[ H, E^{\pm\pm} \right] = \pm E^{\pm\pm} \ , \ \left[ E^{++} , E^{--} \right] = 2 H \ , \ \left[ H , F^\pm \right] = \pm \frac{1}{2} F^\pm \ , \\
	\label{VSPB} && \left[ E^{\pm\pm} , F^\mp \right] = - F^\pm \ , \ \left\lbrace F^\pm , F^\pm \right\rbrace = \pm \frac{1}{2} E^{\pm\pm} \ , \ \left\lbrace F^+ , F^- \right\rbrace = \frac{1}{2} H \ , \\
	\label{VSPC} && \left[ Z, \bar{F}^\pm \right] = \frac{1}{2} F^\pm \ , \ \left\lbrace \bar{F}^\pm , \bar{F}^\pm \right\rbrace = \mp \frac{1}{2} E^{\pm\pm} \ , \\
	\label{VSPD} && \left[ Z , F^\pm \right] = \frac{1}{2} \bar{F}^\pm \ , \ \left[ H , \bar{F}^\pm \right] = \pm \frac{1}{2} \bar{F}^\pm \ , \ \left[ E^{\pm\pm} , \bar{F}^\mp \right] = - \bar{F}^\pm \ , \ \left\lbrace \bar{F}^\pm , F^\mp \right\rbrace = \mp \frac{1}{2} Z \ .
\end{eqnarray}
By denoting with $\mathfrak{g} = \mathfrak{osp}(2|2)$, $\mathfrak{h}= \mathfrak{osp}(1|2)$ and $\mathfrak{k}= \mathfrak{g}/\mathfrak{h}$, we notice that the lines 
\eqref{VSPA} and \eqref{VSPB} represent the subalgebra $\mathfrak{h}$ and, schematically, we have
\begin{equation}\label{VSPE}
	\left[ \mathfrak{h} , \mathfrak{h} \right] \subseteq \mathfrak{h} \ , \ \left[ \mathfrak{h} , \mathfrak{k} \right] \subseteq \mathfrak{k} \ , \ \left[ \mathfrak{k} , \mathfrak{k} \right] \subseteq \mathfrak{h} \ ,
\end{equation}
Then, the subalgebra $\mathfrak{h}$ is \emph{reductive} in $\mathfrak{g}$ and the supercoset $\mathfrak{k}$ is \emph{homogeneous}. 

We can now move on to Maurer-Cartan (MC) forms by translating the commutation relations into MC equations. We denote by $\mathfrak{g}^*, 
\mathfrak{h}^*, \mathfrak{k}^*$ the dual spaces to $\mathfrak{g}, \mathfrak{h}, \mathfrak{k}$: given a vector $\xi$, its dual $\xi^*$ satisfies $\iota_\xi \xi^* =1$. 
The parity inversion is understood, so that $\mathfrak{g}^*$ actually denotes $\Pi \mathfrak{g}^*$. 

The forms dual to the generators are
\begin{equation}\label{VSPF}
	H \leftrightsquigarrow V^0 \ , \ E^{\pm\pm} \leftrightsquigarrow V^{\pm\pm} \ , \ Z \leftrightsquigarrow  U \ , \ F^\pm \leftrightsquigarrow \psi^\pm \ , \ \bar{F}^\pm \leftrightsquigarrow \bar{\psi}^\pm \ ,
\end{equation}
and the MC equations read
\begin{eqnarray}
	\label{VSPG} d V^0 &=& 2 V^{++} \wedge V^{--} + \frac{1}{2} \psi^+ \wedge \psi^- \ , \\
	\label{VSPH} d V^{\pm\pm} &=& \pm V^0 \wedge V^{\pm\pm} \pm \frac{1}{2} \left( \psi^\pm\wedge  \psi^\pm\right) \mp \frac{1}{2} \left( \bar{\psi}^\pm \wedge \bar{\psi}^\pm \right) \ , \\
	\label{VSPI} d \psi^\pm &=& \pm \frac{1}{2} V^0 \wedge \psi^\pm - V^{\pm\pm} \wedge \psi^\mp + \frac{1}{2} U \wedge \bar{\psi}^\pm \ , \\
	\label{VSPJ} d U &=& - \frac{1}{2} \bar{\psi}^+ \wedge \psi^- + \frac{1}{2} \bar{\psi}^- \wedge \psi^+ \ , \\
	\label{VSPK} d \bar{\psi}^\pm &=& \frac{1}{2} U \wedge \psi^\pm \pm \frac{1}{2} V^0 \wedge \bar{\psi}^\pm - V^{\pm\pm} \wedge \bar{\psi}^\mp \ .
\end{eqnarray}
When taking the coset w.r.t. the sub-superalgebra $\mathfrak{osp}(1|2)$, generated by $H, E^{\pm \pm}, F^\pm$ as in \eqref{VSPA} and \eqref{VSPB}, the exterior derivative has to be modified in a covariant derivative as:
\begin{equation}\label{VSPL}
	d \to \nabla_{\mathfrak{k}} = d - A \ ,
\end{equation}
where $A$ is a $3\times 3$ supermatrix whose entries are the $\mathfrak{osp}(1|2)$ MC forms, acting on the vector $(U, \bar{\psi}^+ , \bar{\psi}^-)$. In particular, we 
can rewrite the MC equations in terms of the curvatures of $\mathfrak{osp}(1|2)$: %obtain the operator $\nabla_{\mathfrak{k}}$ with trivial action
\begin{eqnarray}\label{VSPM}
	&&R^0 = 0\,, ~~~~~
	R^{\pm\pm} = \mp \frac{1}{2} \bar{\psi}^\pm \wedge \bar{\psi}^\pm ~~~~~
	\nabla_{\mathfrak{k}}   \psi^\pm = \frac{1}{2} U \wedge \bar{\psi}^\pm \ , \nonumber \\
	&&\nabla_{\mathfrak{k}} U = 0 \,, ~~~~~ \nabla_{\mathfrak{k}} \bar{\psi}^\pm = 0 \ .
\end{eqnarray}
%where $R^0, R^{\pm\pm}, \nabla_{\mathfrak{k}}   \psi^\pm$ are the curvatures of $\mathfrak{osp}(1|2)$. 
This shows that the operator $\nabla_{\mathfrak{k}}$ has a trivial action on any form in $\mathfrak{osp} \left( 2|2 \right) / \mathfrak{osp}(1|2)$, so that they are (covariantly) closed. 

The cohomology of the supercoset $\mathfrak{k}$ is the \emph{equivariant cohomology} (or \emph{relative cohomology}) of $\mathfrak{g}$ relative to $\mathfrak{h}$ \cite{CE}; we first define the \emph{basic} forms %(or \emph{basic} forms) 
of $\mathfrak{k}$ as
\begin{eqnarray}
	\nonumber \left( \Omega^\bullet \left( \mathfrak{g} / \mathfrak{h} \right) \right)^{\mathfrak{h}} \equiv \Omega^\bullet_{basic} \left( \mathfrak{g} \right) &\defeq& \left\lbrace \omega \in \Omega^\bullet \left( \mathfrak{g} / \mathfrak{h} \right) : \mathcal{L}_{\xi} \omega = 0 , \forall \xi \in \mathfrak{h} \right\rbrace  \\
	\label{VSPN} &=& \left\lbrace \omega \in \Omega^\bullet \left( \mathfrak{g} \right) : \iota_\xi \omega = 0 , \mathcal{L}_{\xi} \omega = 0 , \forall \xi \in \mathfrak{h} \right\rbrace \ .
\end{eqnarray}
The cohomology of $\mathfrak{g}$ relative to $\mathfrak{h}$ is then defined as
\begin{equation}\label{VSPO}
	H^\bullet \left( \mathfrak{g}, \mathfrak{h}, \mathbb{R} \right) \equiv H^\bullet_{eq} \left( \mathfrak{g}, \mathfrak{h} \right) \defeq \frac{\left\lbrace \omega \in \left( \Omega^\bullet \left( \mathfrak{g} / \mathfrak{h} \right) \right)^{\mathfrak{h}} : \nabla_{\mathfrak{k}} \omega = 0 \right\rbrace}{\left\lbrace \omega \in \left( \Omega^\bullet \left( \mathfrak{g} / \mathfrak{h} \right) \right)^{\mathfrak{h}} : \exists  \eta \in \left( \Omega^\bullet \left( \mathfrak{g} / \mathfrak{h} \right) \right)^{\mathfrak{h}} \text{s.t. } \omega = \nabla_{\mathfrak{k}} \eta \right\rbrace} \ .
\end{equation}
From now on, we will systematically denote $H^\bullet \left( \mathfrak{g}, \mathfrak{h}, \mathbb{R} \right) \equiv H^\bullet  \left( \mathfrak{g}, \mathfrak{h} \right)$, without specifying that forms are valued in the trivial module $\mathbb{R}$. Going back to the example under examination, we have to look for invariant forms. Since every form in $\mathfrak{k}$ is closed, the invariant ones will automatically generate the relative cohomology. It is easy to demonstrate that among superforms there are no invariants, except for constants:
\begin{equation}\label{VSPP}
	H^p_{super} \left( \mathfrak{g} , \mathfrak{h} \right) = \begin{cases}
		\mathbb{R} \ , \ \text{ if } p = 0 \ , \\
		\left\lbrace 0 \right\rbrace \ , \ \text{ else.}
	\end{cases}
\end{equation}
One has also to consider cohomology classes among integral forms of $\mathfrak{k}$; this is easily done by using the \emph{Berezinian complement} isomorphism introduced in \cite{CCGN2}:
\begin{equation}\label{VSPQ}
	\star : H^{\bullet}_{super} \left( \mathfrak{g} , \mathfrak{h} \right) \overset{\cong}{\underset{}{\longrightarrow}} H^{m-\bullet}_{integral} \left( \mathfrak{g} , \mathfrak{h} \right) \equiv H^{(m-\bullet|n)} \left( \mathfrak{g} , \mathfrak{h} \right) \ ,
\end{equation}
where $m$ and $n$ are the even and odd dimensions of $\mathfrak{k}$, respectively. In this case $m=1, n =2$. This then leads to
\begin{equation}\label{VSPR}
	H^p_{integral} \left( \mathfrak{g} , \mathfrak{h} \right) \equiv H^{(p|2)} \left( \mathfrak{g} , \mathfrak{h} \right) = \begin{cases}
		\mathbb{R} \ , \ \text{ if } p = 1 \ , \\
		\left\lbrace 0 \right\rbrace \ , \ \text{ else.}
	\end{cases}
\end{equation}
The class $\displaystyle \left[ \mathpzc{B}er_{\mathfrak{k}} \right] \in H^1_{integral} \left( \mathfrak{g} , \mathfrak{h} \right)$ corresponds to the Berezinian class of the supercoset $\mathfrak{k}$.

\subsection{The Berezinian $\mathpzc{B}er_{\mathfrak{k}}$: the Explicit Realisation}

We can check that $\displaystyle \left[ \mathpzc{B}er_{\mathfrak{k}} \right]$ is invariant by using an explicit realisation of integral forms (see App. A). 
This realisation was used in \cite{CCGN2} and in many physical contexts (see, e.g., \cite{CG,CG2}).

A representative of the integral form in the class $\displaystyle \left[ \mathpzc{B}er_{\mathfrak{k}} \right]$ is
\begin{equation}\label{AECA}
	\mathpzc{B}er_{\mathfrak{k}} = U \wedge \delta \left( \bar{\psi}^+ \right) \wedge \delta \left( \bar{\psi}^- \right) \ .
\end{equation}
In order to verify that $\mathpzc{B}er_{\mathfrak{k}}$ is invariant w.r.t. the sub-superalgebra $\mathfrak{h}$, we have to verify that
\begin{equation}\label{AECB}
	\iota_\xi \mathpzc{B}er_{\mathfrak{k}} = 0 \ , \ \mathcal{L}_\xi \mathpzc{B}er_{\mathfrak{k}} = 0 \ , \ \forall \xi \in \mathfrak{h} \ .
\end{equation}
While the first condition is trivially satisfied, the second one follows from $\displaystyle d \mathpzc{B}er_{\mathfrak{k}} = 0$ or, explicitly
\begin{eqnarray}
	 d \left[ U \wedge \delta \left( \bar{\psi}^+ \right) \wedge \delta \left( \bar{\psi}^- \right) \right] &=& - \frac12 U \wedge V^0 \wedge \bar{\psi}^+ \wedge \bar{\iota}_+ \delta \left( \bar{\psi}^+ \right) \wedge \delta \left( \bar{\psi}^- \right) + \\
	\nonumber &-& \frac{1}{2} U \wedge \delta \left( \bar{\psi}^+ \right) \wedge V^0 \wedge \bar{\psi}^- \wedge \bar{\iota}_- \delta \left( \bar{\psi}^- \right) = \\
	\label{AECC} &=& \frac{1}{2} U \wedge V^0 \wedge \delta \left( \bar{\psi}^+ \right) \wedge \delta \left( \bar{\psi}^- \right) + \frac{1}{2} U \wedge \delta \left( \bar{\psi}^+ \right) \wedge V^0 \wedge \delta \left( \bar{\psi}^- \right) = 0 \ .\nonumber
\end{eqnarray}
Hence, $\mathcal{L}_{\xi} \mathpzc{B}er_{\mathfrak{k}} = d \iota_\xi \mathpzc{B}er_{\mathfrak{k}} + (-1)^{|\xi|} \iota_\xi d \mathpzc{B}er_{\mathfrak{k}} = 0 , \forall \xi \in \mathfrak{h}$.

With this realisation, we can write the Berezinian $ \mathpzc{B}er_{\mathfrak{g}}$ of the superalgebra ${\mathfrak{g}}$, which represents the top integral form, as
\begin{eqnarray}
\label{newBERA}
\mathpzc{B}er_{\mathfrak{g}} = \mathpzc{B}er_{\mathfrak{k}} \wedge  \mathpzc{B}er_{\mathfrak{h}} = U  \delta \left( \bar{\psi}^+ \right)  \delta \left( \bar{\psi}^- \right)  \wedge    V^0 V^{++} V^{--}  \delta \left({\psi}^+ \right) \delta \left({\psi}^- \right) \ ,
\end{eqnarray}
where the top form $ \mathpzc{B}er_{\mathfrak{k}}$ of the coset \eqref{AECA} multiplies the top form of the Lie subalgebra $ \mathpzc{B}er_{\mathfrak{h}}$. 
$\mathpzc{B}er_{\mathfrak{g}}$ is a top form in $\Omega^{(4|4)} \left( \mathfrak{g} \right)$, then it is closed and not exact. Moreover, notice that, since ${\mathfrak{h}}$ is a Lie subalgebra, the integral form $ \mathpzc{B}er_{\mathfrak{h}}$, i.e., its Berezinian top form, is an element of the cohomology $H^\bullet_{integral}({\mathfrak{h}})$ 
as a consequence of the duality \eqref{VSPQ}. % The factorization into $\mathpzc{B}er_{\mathfrak{k}} \wedge  \mathpzc{B}er_{\mathfrak{h}}$ follows immediately by the spectral sequence discussed next. 

\subsection{The Spectral Sequence}

Let us now focus in defining the pseudoform cohomology by means of Koszul spectral sequences. Firstly, we have to choose the picture number we are working on. In the standard Lie algebra case this point does not clearly arise. In this super-instance, we have to fix \emph{ab initio} in which complex of forms we are constructing the spectral sequence. This is the key point of the derivation of pseudoforms in this framework: we want to show that they naturally arise as integral forms of the sub-structures associated to the given superalgebra, namely, to its sub-superalgebras. In our specific example, this means that we will consider the spectral sequence built at picture number 2. In order to construct page 0 of the spectral sequence we first introduce the filtration (the notation we use is inherited by \cite{Fuks})
\begin{equation}\label{TSSA}
	F^p \Omega^{(q|2)} \left( \mathfrak{g} \right) = \left\lbrace \omega \in \Omega^{(q|2)} \left( \mathfrak{g} \right) : \forall \xi_i \in \mathfrak{h} , \iota_{\xi_{i_1}} \ldots \iota_{\xi_{i_{q+1-p}}} \omega = 0 \right\rbrace \ .
\end{equation}
It is not difficult to verify that
\begin{align}
	\label{TSSB} F^{q+1} \Omega^{(q|2)} \left( \mathfrak{g} \right) = F^{q+2} \Omega^{(q|2)} \left( \mathfrak{g} \right) = \ldots &= F^{q+n} \Omega^{(q|2)} \left( \mathfrak{g} \right) = 0 \ , \ \forall n \in \mathbb{N} \setminus \left\lbrace 0 \right\rbrace , q \in \mathbb{Z} \ , \\
	\label{TSSC} F^{p+1} \Omega^{(q|2)} \left( \mathfrak{g} \right) &\subseteq F^{p} \Omega^{(q|2)} \left( \mathfrak{g} \right) \ , \ \forall p,q \in \mathbb{Z} \ .
\end{align}
In order to verify that \eqref{TSSA} correctly defines a filtration, we have to check that
\begin{equation}\label{TSSD}
	d F^p \Omega^{(q|2)} \left( \mathfrak{g} \right) \subseteq F^p \Omega^{(q+1|2)} \left( \mathfrak{g} \right) \ , \ \forall p,q \in \mathbb{Z} \ .
\end{equation}
Let us verify this for $p=q$ (the generalisation is straightforward). We have
\begin{eqnarray}
	\label{TSSE} F^q \Omega^{(q|2)} \left( \mathfrak{g} \right) &=& \left\lbrace \omega \in \Omega^{(q|2)} \left( \mathfrak{g} \right) : \forall \xi_i \in \mathfrak{h} , \iota_{\xi_i} \omega = 0 \right\rbrace \ , \\
	\label{TSSEA} F^q \Omega^{(q+1|2)} \left( \mathfrak{g} \right) &=& \left\lbrace \omega \in \Omega^{(q+1|2)} \left( \mathfrak{g} \right) : \forall \xi_i \in \mathfrak{h} , \iota_{\xi_{i_1}} \iota_{\xi_{i_2}} \omega = 0 \right\rbrace \ .
\end{eqnarray}
We have that $d \omega \in F^q \Omega^{(q+1|2)} \left( \mathfrak{g} \right)$ iff $\iota_{\xi_1} \iota_{\xi_2} d\omega = 0, \forall \xi_1 , \xi_2 \in \mathfrak{h}$. In particular, we have
\begin{equation}\label{TSSF}
	\iota_{\xi_1} \iota_{\xi_2} d\omega = \iota_{\xi_1} \iota_{\xi_2} d\omega + \left( -1 \right)^{|\xi_2|+1} \iota_{\xi_1} d \iota_{\xi_2} \omega = $$ $$ = \iota_{\xi_1} \mathcal{L}_{\xi_2} \omega = \iota_{\xi_1} \mathcal{L}_{\xi_2} \omega + \left( -1 \right)^{|\xi_1||\xi_2|+1} \mathcal{L}_{\xi_2} \iota_{\xi_1} \omega = \iota_{\left[ \xi_1 , \xi_2 \right]} \omega = 0 \ ,
\end{equation}
where we systematically added trivial terms and used the definitions \eqref{TSSE} and \eqref{TSSEA}. The extension to any $p$ leads to the same type of manipulations. Hence we verified that \eqref{TSSA} correctly defines a filtration on $\Omega^{(\bullet|2)} \left( \mathfrak{g} \right)$.

There are major differences between conventional Lie algebras and Lie superalgebras:
for the former we have $q \in \left\lbrace 0 , 1 , \ldots , \text{dim} \, \mathfrak{g} \right\rbrace$, for the latter $q \in \mathbb{Z}$. This is a consequence of the fact that the complex of superforms is unbounded from above and the complex of integral forms is unbounded from below. 
Since the pseudoforms induced from the filtration \eqref{TSSA} arise as 
\begin{equation}\label{TSSFA}
	\Omega_{integral}^\bullet \left( \mathfrak{k} \right) \otimes \Omega_{super}^\bullet \left( \mathfrak{h} \right) \ ,
\end{equation}
these complexes are unbounded both from above and from below. 

In addition, for Lie algebras $\mathfrak{g}$, one always has
	\begin{equation}\label{TSSG}
		F^p \Omega^q  \left( \mathfrak{g} \right) = \Omega^q \left( \mathfrak{g} \right) \ , \ \forall p \leq 0 \ ,
	\end{equation}
since the contraction operator $\iota_\xi$ is odd for any $\xi \in \mathfrak{g}$. On the contrary, 
for the integral forms complex of superalgebras \eqref{TSSG} does not hold.

The spaces of pseudoforms of the form \eqref{TSSFA}, for any form number $q$, are defined as
\begin{eqnarray}
\label{AN_A}
\Omega^{(q|2)}(\mathfrak{g}) = \bigoplus_{r \leq b(\mathfrak{k})} \Omega^{(r|2)}(\mathfrak{k}) \otimes \Omega^{(q-r|0)}(\mathfrak{h}) \ ,
\end{eqnarray}
where $b(\mathfrak{k})$ is the even dimension of the supercoset $\mathfrak{k}$. The sum starts with 
$ \Omega^{(b(\mathfrak{k})|2)}(\mathfrak{k}) \otimes \Omega^{(q-b(\mathfrak{k})|0)}(\mathfrak{h})$,
where $\Omega^{(b(\mathfrak{k})|2)}(\mathfrak{k})$ is the Berezinian bundle of $\mathfrak{k}$. %The factors $\Big(\bigoplus_{s=1}^{q-b(\mathfrak{k})} \Omega^{(s|0)}(\mathfrak{h})\Big)$ take care of form numbers $q > b(\mathfrak{k})$. 
It is convenient to use the following notation for the spaces of the filtration \eqref{TSSA}
\begin{equation}\label{TSSH}
	F^{p} \Omega^{(q|2)}\left( \mathfrak{g} \right) \equiv \Omega^{(q|2)}_{q-p} \left( \mathfrak{g} \right) \defeq \bigoplus_{i=0}^{q-p}\Omega^{(q-i|2)}(\mathfrak{k}) \otimes  \Omega^{(i|0)}(\mathfrak{h}) \ , 
	%\left\lbrace \omega \in \Omega^{(r|2)} \left( \mathfrak{g} \right) : \forall \xi_l \in \mathfrak{h} , \iota_{\xi_1} \ldots \iota_{\xi_{i+1}} \omega = 0 \right\rbrace \ ,
\end{equation}
i.e., the space $(q|2)$-forms depending \emph{at most} on $q-p$-superforms in 
$\mathfrak{h}^*$. For example, 
\begin{eqnarray}
\label{AN_B}
\Omega^{(q|2)}_{1} \left( \mathfrak{g} \right) &=& 
\Omega^{(q|2)} \left( \mathfrak{k} \right) \oplus \Big(\Omega^{(q-1|2)} \left( \mathfrak{k} \right) \otimes \Omega^{(1|0)} \left( \mathfrak{h}\right) \Big) \ , \nonumber \\
\Omega^{(q|2)}_{2} \left( \mathfrak{g} \right) &=& \Omega^{(q|2)} \left( \mathfrak{k} \right) \oplus 
\Big(\Omega^{(q-1|2)} \left( \mathfrak{k} \right) \otimes \Omega^{(1|0)} \left( \mathfrak{h}\right)\Big)
 \oplus \Big(\Omega^{(q-2|2)} \left( \mathfrak{k} \right) \otimes \Omega^{(2|0)} \left( \mathfrak{h}\right)\Big) \ . \nonumber 
\end{eqnarray}
In the case $\text{dim} \mathfrak{k} = (1|2)$, namely, when $ \Omega^{(q|2)} \left( \mathfrak{k} \right) = \left\lbrace 0 \right\rbrace$ 
for $q >1$, the definition in \eqref{TSSA} implies 
\begin{eqnarray}\label{TSSI}
	&&F^p \Omega^{(q|2)} \left( \mathfrak{g} \right) = 0 \ , \ \forall p \geq 2 \ , \ \forall q \geq 1 \ , \nonumber \\
	&&F^1 \Omega^{(q|2)} \left( \mathfrak{g} \right) = \Omega_{q-1}^{(q|2)}\left( \mathfrak{g} \right) =\Omega^{(1|2)} \left( \mathfrak{k} \right) \otimes \Omega^{(q-1|0)} \left( \mathfrak{h} \right) \ , \ \forall q \geq 1 \ ,
\end{eqnarray}
and we can conveniently display the relevant spaces in Table \ref{TableTSSA}.
\begin{table}[ht!]
\centering
\begin{tabular}{cccccccccc}
\multicolumn{1}{c|}{$\ldots$} & \multicolumn{1}{c|}{$\ldots$}& \multicolumn{1}{c|}{$\ldots$} & \multicolumn{1}{c|}{$\ldots$} & \multicolumn{1}{c|}{$\vdots$} & \multicolumn{1}{c|}{$\ldots$}& \multicolumn{1}{c|}{$\ldots$} & \multicolumn{1}{c|}{$\ldots$} & \multicolumn{1}{c|}{$\ldots$} & $\ldots$ 
%\\ 
%\cline{1-4} \cline{6-10} 
%\multicolumn{1}{c|}{$\ldots$} & \multicolumn{1}{c|}{0}& \multicolumn{1}{c|}{0}& \multicolumn{1}{c|}{0}& \multicolumn{1}{c|}{3}& \multicolumn{1}{c|}{0}& \multicolumn{1}{c|}{0}                                                                           & \multicolumn{1}{c|}{0} & \multicolumn{1}{c|}{$\displaystyle \Omega^{(3|2)} \left( \mathfrak{k} \right)$} & $\ldots$ 
%\\ 
%\cline{1-4} \cline{6-10} \multicolumn{1}{c|}{$\ldots$} & \multicolumn{1}{c|}{0} & \multicolumn{1}{c|}{0} & \multicolumn{1}{c|}{0}  & \multicolumn{1}{c|}{2}        & \multicolumn{1}{c|}{0}                                                                           & \multicolumn{1}{c|}{0}  & \multicolumn{1}{c|}{$\displaystyle \Omega^{(2|2)} \left( \mathfrak{k} \right)$}
%& \multicolumn{1}{c|}{$\displaystyle \Omega^{(3|2)}_{~1} \left( \mathfrak{g} \right)$} & $\ldots$ 
\\ 
\cline{1-4} \cline{6-10} 
\multicolumn{1}{c|}{$\ldots$} & \multicolumn{1}{c|}{0} & \multicolumn{1}{c|}{0}  & \multicolumn{1}{c|}{0}  & \multicolumn{1}{c|}{1}        & \multicolumn{1}{c|}{0}                                                                           & \multicolumn{1}{c|}{$\displaystyle \Omega^{(1|2)} \left( \mathfrak{k} \right)$}                  & \multicolumn{1}{c|}{$\displaystyle \Omega^{(2|2)}_{1} \left( \mathfrak{g} \right)$} & \multicolumn{1}{c|}{$\displaystyle \Omega^{(3|2)}_{2} \left( \mathfrak{g} \right)$} & $\ldots$ 
\\ \cline{1-4} \cline{6-10} 
\multicolumn{1}{c|}{$\ldots$} & \multicolumn{1}{c|}{0}                                                           & \multicolumn{1}{c|}{0}                                                                            & \multicolumn{1}{c|}{0}                                                                            & \multicolumn{1}{c|}{0}        & \multicolumn{1}{c|}{$\displaystyle \Omega^{(0|2)} \left( \mathfrak{k} \right)$}                  & \multicolumn{1}{c|}{$\displaystyle \Omega^{(1|2)}_{~1} \left( \mathfrak{g} \right)$} & \multicolumn{1}{c|}{$\displaystyle \Omega^{(2|2)}_{~2} \left( \mathfrak{g} \right)$} & \multicolumn{1}{c|}{$\displaystyle \Omega^{(3|2)}_{~3} \left( \mathfrak{g} \right)$} & $\ldots$ \\ \cline{1-4} \cline{6-10} 
$\ldots$                      & -3                                                                               & -2                                                                                                & -1                                                                                                &                               & 0                                                                                                & 1                                                                                                & 2                                                                                                & 3                                                                                                & $\ldots$ \\ \cline{1-4} \cline{6-10} 
\multicolumn{1}{c|}{$\ldots$} & \multicolumn{1}{c|}{0}                                                           & \multicolumn{1}{c|}{0}                                                                            & \multicolumn{1}{c|}{$\displaystyle \Omega^{(-1|2)} \left( \mathfrak{k} \right)$}                  & \multicolumn{1}{c|}{-1}       & \multicolumn{1}{c|}{$\displaystyle \Omega^{(0|2)}_{~1} \left( \mathfrak{g} \right)$} & \multicolumn{1}{c|}{$\displaystyle \Omega^{(1|2)}_{~2} \left( \mathfrak{g} \right)$} & \multicolumn{1}{c|}{$\displaystyle \Omega^{(2|2)}_{~3} \left( \mathfrak{k} \right)$} & \multicolumn{1}{c|}{$\displaystyle \Omega^{(3|2)}_{~4} \left( \mathfrak{g} \right)$} & $\ldots$ \\ \cline{1-4} \cline{6-10} 
\multicolumn{1}{c|}{$\ldots$} & \multicolumn{1}{c|}{0}                                                           & \multicolumn{1}{c|}{$\displaystyle \Omega^{(-2|2)} \left( \mathfrak{k} \right)$}                  & \multicolumn{1}{c|}{$\displaystyle \Omega^{(-1|2)}_{~1} \left( \mathfrak{g} \right)$} & \multicolumn{1}{c|}{-2}       & \multicolumn{1}{c|}{$\displaystyle \Omega^{(0|2)}_{~2} \left( \mathfrak{g} \right)$} & \multicolumn{1}{c|}{$\displaystyle \Omega^{(1|2)}_{~3} \left( \mathfrak{g} \right)$} & \multicolumn{1}{c|}{$\displaystyle \Omega^{(2|2)}_{~4} \left( \mathfrak{g} \right)$} & \multicolumn{1}{c|}{$\displaystyle \Omega^{(3|2)}_{~5} \left( \mathfrak{g} \right)$} & $\ldots$ \\ \cline{1-4} \cline{6-10} 
\multicolumn{1}{c|}{$\ldots$} & \multicolumn{1}{c|}{$\displaystyle \Omega^{(-3|2)} \left( \mathfrak{k} \right)$} & \multicolumn{1}{c|}{$\displaystyle \Omega^{(-2|2)}_{~1} \left( \mathfrak{g} \right)$} & \multicolumn{1}{c|}{$\displaystyle \Omega^{(-1|2)}_{~2} \left( \mathfrak{g} \right)$} & \multicolumn{1}{c|}{-3}       & \multicolumn{1}{c|}{$\displaystyle \Omega^{(0|2)}_{~3} \left( \mathfrak{g} \right)$} & \multicolumn{1}{c|}{$\displaystyle \Omega^{(1|2)}_{~4} \left( \mathfrak{g} \right)$} & \multicolumn{1}{c|}{$\displaystyle \Omega^{(2|2)}_{~5} \left( \mathfrak{g} \right)$} & \multicolumn{1}{c|}{$\displaystyle \Omega^{(3|2)}_{~6} \left( \mathfrak{g} \right)$} & $\ldots$ \\ \cline{1-4} \cline{6-10} 
\multicolumn{1}{c|}{$\ldots$} & \multicolumn{1}{c|}{$\ldots$}                                                    & \multicolumn{1}{c|}{$\ldots$}                                                                     & \multicolumn{1}{c|}{$\ldots$}                                                                     & \multicolumn{1}{c|}{$\vdots$} & \multicolumn{1}{c|}{$\ldots$}                                                                    & \multicolumn{1}{c|}{$\ldots$}                                                                    & \multicolumn{1}{c|}{$\ldots$}                                                                    & \multicolumn{1}{c|}{$\ldots$}                                                                    & $\ldots$
\end{tabular}
\caption{Filtration defined in \eqref{TSSA}. The integers $q$ and $p$ are spanned on the horizontal and vertical axes, respectively.}\label{TableTSSA}
\end{table}
 
We can now define page 0 of the spectral sequence for the pseudoforms of $\mathfrak{g}$ associated to the filtration \eqref{TSSA} as
\begin{equation}\label{TSSJ}
	E_0^{m,n} \defeq F^m \Omega^{(m+n|2)} \left( \mathfrak{g} \right) / F^{m+1} \Omega^{(m+n|2)} \left( \mathfrak{g} \right) \ .
\end{equation}
In Table \ref{TableTSSB} we collect the whole page at picture number 2 for the example under examination.
\vskip .3cm
\begin{table}[ht!]
\centering
\begin{tabular}{cccccc}
%$\ldots$                      & -1                            &
 \multicolumn{1}{c}~                               
&  \multicolumn{1}{c}0                                                                  
&  \multicolumn{1}{c}1                                                                                                                     
&  \multicolumn{1}{c}2                                                                                                                     
&  \multicolumn{1}{c}3                                                                                                                     
& $\ldots$ 
\\ \cline{2-6} %\cline{4-8} 
%\multicolumn{1}{c|}{$\ldots$} & \multicolumn{1}{c|}{$\ldots$} & 
\multicolumn{1}{c|}{$\vdots$} 
& \multicolumn{1}{c|}{$\ldots$}                                      
& \multicolumn{1}{c|}{$\ldots$}                                                                                         
& \multicolumn{1}{c|}{$\ldots$}                                                                                         
& \multicolumn{1}{c|}{$\ldots$}                                                                                         
& $\ldots$ 
\\ \cline{2-6} %\cline{4-8} 
%\multicolumn{1}{c|}{$\ldots$} & \multicolumn{1}{c|}{0}        &
\multicolumn{1}{c|}{2}        
& \multicolumn{1}{c|}{0}                                             
& \multicolumn{1}{c|}{0}                                                                                                
& \multicolumn{1}{c|}{0}                                                                                                
& \multicolumn{1}{c|}{0}                                                                                                
& $\ldots$ 
\\ \cline{2-6} %\cline{4-8} 
%\multicolumn{1}{c|}{$\ldots$} & \multicolumn{1}{c|}{0}      &  
\multicolumn{1}{c|}{1}        
& \multicolumn{1}{c|}{$\Omega^{(1|2)} \left( \mathfrak{k} \right)$}  
& \multicolumn{1}{c|}{$\Omega^{(1|2)} \left( \mathfrak{k} \right) \otimes \Omega^{(1|0)} \left( \mathfrak{h} \right)$}  
& \multicolumn{1}{c|}{$\Omega^{(1|2)} \left( \mathfrak{k} \right) \otimes \Omega^{(2|0)} \left( \mathfrak{h} \right)$}  
& \multicolumn{1}{c|}{$\Omega^{(1|2)} \left( \mathfrak{k} \right) \otimes \Omega^{(3|0)} \left( \mathfrak{h} \right)$}  
& $\ldots$ 
\\ \cline{2-6} %\cline{4-8} 
%\multicolumn{1}{c|}{$\ldots$} & \multicolumn{1}{c|}{0}        & 
\multicolumn{1}{c|}{0}        
& \multicolumn{1}{c|}{$\Omega^{(0|2)} \left( \mathfrak{k} \right)$}  
& \multicolumn{1}{c|}{$\Omega^{(0|2)} \left( \mathfrak{k} \right) \otimes \Omega^{(1|0)} \left( \mathfrak{h} \right)$}  
& \multicolumn{1}{c|}{$\Omega^{(0|2)} \left( \mathfrak{k} \right) \otimes \Omega^{(2|0)} \left( \mathfrak{h} \right)$}  
& \multicolumn{1}{c|}{$\Omega^{(0|2)} \left( \mathfrak{k} \right) \otimes \Omega^{(3|0)} \left( \mathfrak{h} \right)$}  
& $\ldots$ 
\\ \cline{2-6} %\cline{4-8} 
%\multicolumn{1}{c|}{$\ldots$} & \multicolumn{1}{c|}{0}        & 
\multicolumn{1}{c|}{-1}       
& \multicolumn{1}{c|}{$\Omega^{(-1|2)} \left( \mathfrak{k} \right)$} 
& \multicolumn{1}{c|}{$\Omega^{(-1|2)} \left( \mathfrak{k} \right) \otimes \Omega^{(1|0)} \left( \mathfrak{h} \right)$} 
& \multicolumn{1}{c|}{$\Omega^{(-1|2)} \left( \mathfrak{k} \right) \otimes \Omega^{(2|0)} \left( \mathfrak{h} \right)$} 
& \multicolumn{1}{c|}{$\Omega^{(-1|2)} \left( \mathfrak{k} \right) \otimes \Omega^{(3|0)} \left( \mathfrak{h} \right)$} 
& $\ldots$ 
\\ \cline{2-6} %\cline{4-8} 
%\multicolumn{1}{c|}{$\ldots$} & \multicolumn{1}{c|}{0}        & 
\multicolumn{1}{c|}{-2}       
& \multicolumn{1}{c|}{$\Omega^{(-2|2)} \left( \mathfrak{k} \right)$} 
& \multicolumn{1}{c|}{$\Omega^{(-2|2)} \left( \mathfrak{k} \right) \otimes \Omega^{(1|0)} \left( \mathfrak{h} \right)$} 
& \multicolumn{1}{c|}{$\Omega^{(-2|2)} \left( \mathfrak{k} \right) \otimes \Omega^{(2|0)} \left( \mathfrak{h} \right)$} 
& \multicolumn{1}{c|}{$\Omega^{(-2|2)} \left( \mathfrak{k} \right) \otimes \Omega^{(3|0)} \left( \mathfrak{h} \right)$} 
& $\ldots$ 
\\ \cline{2-6} %\cline{4-8} 
%\multicolumn{1}{c|}{$\ldots$} & \multicolumn{1}{c|}{0}        &
\multicolumn{1}{c|}{-3}       
& \multicolumn{1}{c|}{$\Omega^{(-3|2)} \left( \mathfrak{k} \right)$} 
& \multicolumn{1}{c|}{$\Omega^{(-3|2)} \left( \mathfrak{k} \right) \otimes \Omega^{(1|0)} \left( \mathfrak{h} \right)$} 
& \multicolumn{1}{c|}{$\Omega^{(-3|2)} \left( \mathfrak{k} \right) \otimes \Omega^{(2|0)} \left( \mathfrak{h} \right)$} 
& \multicolumn{1}{c|}{$\Omega^{(-3|2)} \left( \mathfrak{k} \right) \otimes \Omega^{(3|0)} \left( \mathfrak{h} \right)$} 
& $\ldots$ 
\\ \cline{2-6} %\cline{4-8} 
%\multicolumn{1}{c|}{$\ldots$} & \multicolumn{1}{c|}{$\ldots$} & 
\multicolumn{1}{c|}{$\vdots$} 
& \multicolumn{1}{c|}{$\ldots$}                                      
& \multicolumn{1}{c|}{$\ldots$}                                                                                         
& \multicolumn{1}{c|}{$\ldots$}                                                                                         
& \multicolumn{1}{c|}{$\ldots$}                                                                                         
& $\ldots$
\end{tabular}
\caption{$E_0^{m,n}$ as defined in \eqref{TSSJ}. The integers $n$ and $m$ are spanned on the horizontal and vertical axes, respectively.}\label{TableTSSB}
\end{table}

%Notice that differently from purely bosonic Lie algebras, the page represented in Table \ref{TableTSSB} extends to negative form numbers, as a consequence of the presence of integral forms. 
In order to proceed with the construction of the spectral sequence we have to define the differentials to move from one page to the other. First of all, from the MC equations (\ref{VSPG}), (\ref{VSPH}), (\ref{VSPI}), (\ref{VSPJ}) and (\ref{VSPK}), we can schematically write the CE differential for $\mathfrak{g} = \mathfrak{osp} (2|2)$ and $\mathfrak{h} = \mathfrak{osp} (1|2)$ as
\begin{equation}\label{TSSK}
	d = V_{\mathfrak{h}} V_{\mathfrak{h}} \iota_{\mathfrak{h}} + V_{\mathfrak{k}} V_{\mathfrak{k}} \iota_{\mathfrak{h}} + V_{\mathfrak{k}} V_{\mathfrak{h}} \iota_{\mathfrak{k}} \ .
\end{equation}
The first differential $d_0$ of the spectral sequence is the \emph{horizontal differential} induced by $d$, acting on objects in Table \ref{TableTSSB} as
\begin{eqnarray}\label{TSSL}
	d_0 : E_0^{m,n} &\longrightarrow& E_0^{m,n+1} \nonumber \\
	\omega &\mapsto& d_0 \omega \defeq \left( V_{\mathfrak{k}} V_{\mathfrak{h}} \iota_{\mathfrak{k}} + V_{\mathfrak{h}} V_{\mathfrak{h}} \iota_{\mathfrak{h}} \right) \omega \ \ .
\end{eqnarray}
Page 1 of the spectral sequence is defined as the cohomology of page 0 with respect to  the differential $d_0$:
\begin{equation}\label{TSSM}
	E_1^{m,n} \defeq H \left( E_0^{m,n} , d_0 \right) \ .
\end{equation}
We can explicitly calculate page 1 in the case under consideration: given $d_0$ as in \eqref{TSSL}, we can treat the two factors 
in the tensor product of $E_0^{m,n}$ separately. In particular, we have
\begin{equation}\label{TSSN}
	H \left( \Omega^{(m|2)} \left( \mathfrak{k} \right) \otimes \Omega^{(n|0)} \left( \mathfrak{h} \right) , V_{\mathfrak{h}} V_{\mathfrak{h}} \iota_{\mathfrak{h}} \right) = \Omega^{(m|2)} \left( \mathfrak{k} \right) \otimes H^{(n|0)} \left( \mathfrak{h} \right) \ ,
\end{equation}
since this part of the differential selects the cohomology of superforms in the Lie sub-superalgebra $\mathfrak{h}$ and it does not act on the first factor. 

For the other part of the differential \eqref{TSSL}, we can express its action as
\begin{equation}\label{TSSO}
	V_{\mathfrak{k}} V_{\mathfrak{h}} \iota_{\mathfrak{k}} \left( \omega_{\mathfrak{k}} \otimes \omega_{\mathfrak{h}} \right) = \left( V_{\mathfrak{k}} V_{\mathfrak{h}} \iota_{\mathfrak{k}} \otimes 1 \right) \left( \omega_{\mathfrak{k}} \otimes \omega_{\mathfrak{h}} \right) = 0 \ \iff \ \left( \mathcal{L}_{\mathfrak{h}} \otimes 1 \right) \left( \omega_{\mathfrak{k}} \otimes \omega_{\mathfrak{h}} \right) = 0 \ ,
\end{equation}
on the form $\omega_{\mathfrak{k}} \otimes \omega_{\mathfrak{h}} \in \Omega^{(m|2)} \left( \mathfrak{k} \right)  \otimes \Omega^{(n|0)} \left( \mathfrak{h} \right)$,
where with $\mathcal{L}_{\mathfrak{h}}$ we formally denote the Lie derivative along any vector in $\mathfrak{h}$. The double implication follows from the equivalence $d_0 \omega_{\mathfrak{k}} = d \omega_{\mathfrak{k}}$, which follows 
%
%\eqref{TSSO} is easily verified, since we can see from \eqref{TSSK} that in the case under examination we have
%\begin{equation}\label{TSSP}
%	V_{\mathfrak{k}} V_{\mathfrak{h}} \iota_{\mathfrak{k}} \otimes 1 \equiv d \otimes 1 \ ,
%\end{equation}
since the other two terms of \eqref{TSSK} vanish.  
This means that
\begin{equation}\label{TSSQ}
	H \left( \Omega^{(m|2)} \left( \mathfrak{k} \right) \otimes \Omega^{(n|0)} \left( \mathfrak{h} \right) , V_{\mathfrak{k}} V_{\mathfrak{h}} \iota_{\mathfrak{k}} \right) = \left( \Omega^{(m|2)} \left( \mathfrak{k} \right) \right)^{\mathfrak{h}} \otimes \Omega^{(n|0)} \left( \mathfrak{h} \right) \ ,
\end{equation}
i.e., we have to take $\mathfrak{h}$-\emph{invariant} forms in $\Omega^{(m|2)} \left( \mathfrak{k} \right)$. Finally, we get
\begin{equation}\label{TSSR}
	E_1^{m,n} = \left( \Omega^{(m|2)} \left( \mathfrak{k} \right) \right)^{\mathfrak{h}} \otimes H^{(n|0)} \left( \mathfrak{h} \right) \ .
\end{equation}
%Using reductivity hypothesis of the sub-superalgebra $\mathfrak{h} \subset \mathfrak{g}$, 
The following page of the spectral sequence is defined as
\begin{eqnarray}\label{TSSS}
	E_2^{m,n} \defeq H \left( E_1^{m,n} , d_1 \right) \ , 
\end{eqnarray}
where  $d_1$ is the \emph{vertical operator} $\displaystyle d_1 : E_1^{m,n} \to  E_1^{m+1,n}$, which increases the form number in the coset direction by one. In this case $d_1$ is trivial, as one can readily see from \eqref{TSSK}.

In a general setting, the CE differential reads
\begin{equation}\label{TSST}
	d = V_{\mathfrak{h}} V_{\mathfrak{h}} \iota_{\mathfrak{h}} + V_{\mathfrak{k}} V_{\mathfrak{k}} \iota_{\mathfrak{h}} + V_{\mathfrak{k}} V_{\mathfrak{h}} \iota_{\mathfrak{k}} + V_{\mathfrak{k}} V_{\mathfrak{k}} \iota_{\mathfrak{k}} + V_{\mathfrak{k}} V_{\mathfrak{h}} \iota_{\mathfrak{h}} \ ,
\end{equation}
so that $d_1$ in general reads
\begin{equation}\label{TSSU}
	d_1 = V_{\mathfrak{k}} V_{\mathfrak{k}} \iota_{\mathfrak{k}} + V_{\mathfrak{k}} V_{\mathfrak{h}} \iota_{\mathfrak{h}} \ .
\end{equation}
The first term amounts to the CE differential which is used in order to calculate the relative cohomology, defined in \eqref{VSPL}, which in our case is trivial as shown in \eqref{VSPM}; the second term is zero for reductive sub-superalgebras in the ambient superalgebra, which is true in the case under examination. Then we get
\begin{equation}\label{TSSV}
	\left( \Omega^{(p|2)} \left( \mathfrak{k} \right) \right)^{\mathfrak{h}} = H^{(p|2)} \left( \mathfrak{g} , \mathfrak{h} \right) \ \ \implies \ \ E_2^{m,n} = H^{(m|2)} \left( \mathfrak{g} , \mathfrak{h} \right) \otimes H^{(n|0)} \left( \mathfrak{h} \right) = E_1^{m,n} \ .
\end{equation}
Page 2 for the guiding example is reported in Table \ref{TableTSSC}.
\begin{table}[ht!]
\centering
\begin{tabular}{ccccccccc}
\multicolumn{1}{c|}{$\ldots$} & \multicolumn{1}{c|}{$\ldots$} & \multicolumn{1}{c|}{$\vdots$} & \multicolumn{1}{c|}{$\ldots$}                                                            & \multicolumn{1}{c|}{$\ldots$} & \multicolumn{1}{c|}{$\ldots$} & \multicolumn{1}{c|}{$\ldots$}                                                                                                        & \multicolumn{1}{c|}{$\ldots$} & $\ldots$ \\ \cline{1-2} \cline{4-9} 
\multicolumn{1}{c|}{$\ldots$} & \multicolumn{1}{c|}{0}        & \multicolumn{1}{c|}{2}        & \multicolumn{1}{c|}{0}                                                                   & \multicolumn{1}{c|}{0}        & \multicolumn{1}{c|}{0}        & \multicolumn{1}{c|}{0}                                                                                                               & \multicolumn{1}{c|}{0}        & $\ldots$ \\ \cline{1-2} \cline{4-9} 
\multicolumn{1}{c|}{$\ldots$} & \multicolumn{1}{c|}{0}        & \multicolumn{1}{c|}{1}        & \multicolumn{1}{c|}{$H^{(1|2)} \left( \mathfrak{g} , \mathfrak{h} \right)$} & \multicolumn{1}{c|}{0}        & \multicolumn{1}{c|}{0}        & \multicolumn{1}{c|}{$H^{(1|2)} \left( \mathfrak{g} , \mathfrak{h} \right) \otimes H^{(3|0)} \left( \mathfrak{h} \right)$} & \multicolumn{1}{c|}{0}        & $\ldots$ \\ \cline{1-2} \cline{4-9} 
\multicolumn{1}{c|}{$\ldots$} & \multicolumn{1}{c|}{0}        & \multicolumn{1}{c|}{0}        & \multicolumn{1}{c|}{0}                                                                   & \multicolumn{1}{c|}{0}        & \multicolumn{1}{c|}{0}        & \multicolumn{1}{c|}{0}                                                                                                               & \multicolumn{1}{c|}{0}        & $\ldots$ \\ \cline{1-2} \cline{4-9} 
$\ldots$                      & -1                            &                               & 0                                                                                        & 1                             & 2                             & 3                                                                                                                                    & 4                             & $\ldots$ \\ \cline{1-2} \cline{4-9} 
\multicolumn{1}{c|}{$\ldots$} & \multicolumn{1}{c|}{0}        & \multicolumn{1}{c|}{-1}       & \multicolumn{1}{c|}{0}                                                                   & \multicolumn{1}{c|}{0}        & \multicolumn{1}{c|}{0}        & \multicolumn{1}{c|}{0}                                                                                                               & \multicolumn{1}{c|}{0}        & $\ldots$ \\ \cline{1-2} \cline{4-9} 
\multicolumn{1}{c|}{$\ldots$} & \multicolumn{1}{c|}{$\ldots$} & \multicolumn{1}{c|}{$\vdots$} & \multicolumn{1}{c|}{$\ldots$}                                                            & \multicolumn{1}{c|}{$\ldots$} & \multicolumn{1}{c|}{$\ldots$} & \multicolumn{1}{c|}{$\ldots$}                                                                                                        & \multicolumn{1}{c|}{$\ldots$} & $\ldots$
\end{tabular}
\caption{$E_2^{m,n}$ as obtained in \eqref{TSSV}. The integers $n$ and $m$ are spanned on the horizontal and vertical axes, respectively.}\label{TableTSSC}
\end{table}

We should now proceed with the construction of the higher pages until the spectral sequence converges. This is done by considering the differentials $d_s$:
\begin{equation}\label{TSSW}
	d_s: E_s^{m,n} \to E_s^{m+s,n-s+1} \ ,
\end{equation}
induced by the Koszul differential $d$. In particular, we notice that $d_s$ moves vertically by $s$ and horizontally by $1-s$. In the example under examination, all the higher differentials $d_s$ are trivial:
\begin{equation}\label{TSSX}
	d_s = 0 \ , \ \forall s \geq 2 \ .
\end{equation}
This means that $E_1^{m,n} = E_2^{m,n} = E_3^{m,n} = \ldots = E_\infty^{m,n}$, and the non-trivial cohomology spaces are 
\begin{equation}\label{TSSY}
	H^{1|2} \left( \mathfrak{g} \right) = \mathbb{R} \ , \ H^{4|2} \left( \mathfrak{g} \right) = \Pi \mathbb{R} \ ,
\end{equation}
which are pseudoform cohomology spaces, i.e., with non-zero and non-maximal picture number.

We have shown through spectral sequences that pseudoforms can be constructed in an algebraic context. They are related to sub-structures (in the specific example, we refer to the non-trivial sub-superalgebra 
$\mathfrak{osp}(1|2)$), as in the case of sub-supermanifolds with non-trivial odd codimension (see \cite{Witten}).
 
Second, by considering the case of $\mathfrak{osp} \left( 2|2 \right) $, the cohomology according to  \cite{Fuks} is  
 \begin{equation}\label{TSSZ}
	%H^p_{super} \left( \mathfrak{osp} \left( 2|2 \right) \right) \equiv 
	H^{(p)} \left( \mathfrak{osp} \left( 2|2 \right) \right) = H^p \left( \mathfrak{sp} \left( 2 \right) \right) \ ,
\end{equation}
which in our case amount to superform cohomology only. \eqref{TSSZ} is incomplete since it does not take into account the cohomology of pseudoforms and integral forms. Further, one can notice that, 
according to \eqref{TSSZ}, the sub-algebra $\mathfrak{so} (2)$ plays no role in $H^p_{super} \left( \mathfrak{osp} \left( 2|2 \right) \right)$, then one might 
expect that this emerges by considering the rest of the cohomology: for that we have to complete the cohomology with pseudoforms and integral forms. 

% The point is actually deeper than this: since it is algebraic, the Lie superalgebras cohomology is related to the invariants of the given superalgebras or, in other words, to their rank. The classification restricted to the complex of superforms does not consider all the invariants of the given superalgebra, but only a portion of them. In the case described above, this is reflected in
%
%i.e., the rank carried by the abelian sub-algebra $\mathfrak{so} (2)$ is not represented among superforms. This may lead to think \emph{a priori} that some invariants (or equivalently the rank corresponding to them) may be lost, but what we actually demonstrated in the previous case is that these invariants are instead \virgolette spread" on the whole complex of forms, including pseudoforms and integral forms. In the following section we will further discuss this point by using a generalised Poincar\'e polynomial that takes into account both the form number and the picture number.

The pseudoforms we obtained in \eqref{TSSY} are not the only ones for $\mathfrak{osp}(2|2)$. The additional 
pseudoforms arise from a new filtration which is inequivalent to \eqref{TSSA}. In \cite{Witten}, superforms and integral forms were introduced in the context of Clifford-Weyl algebras; there, it is shown that while for conventional manifolds the irreducible modules are all isomorphic (there is a single complex of forms), for supermanifolds the modules constructed 
from a state annihilated by all the contractions and the modules constructed from a state annihilated by form multiplication are inequivalent, and they are identified with superforms and integral forms, respectively. In our algebraic setting, this suggests a different way in which we can select a filtration with respect to the one in \eqref{TSSA}, i.e., %(again, we stick to the specific example under examination):
\begin{equation}\label{TSSZA}
	\tilde{F}^p \Omega^{(q|2)} \left( \mathfrak{g} \right) = \left\lbrace \omega \in \Omega^{(q|2)} \left( \mathfrak{g} \right) : \forall \xi^*_i \in \mathfrak{h}^* , 
	\xi^*_{i_1} \wedge \ldots \wedge \xi^*_{i_{q+1-p}} \wedge \omega = 0 \right\rbrace \ .
\end{equation}
From \eqref{TSSZA} one can easily verify that
\begin{align}
	\label{TSSZB} \tilde{F}^{q+1} \Omega^{(q|2)} \left( \mathfrak{g} \right) = \tilde{F}^{q+2} \Omega^{(q|2)} \left( \mathfrak{g} \right) = \ldots &= \tilde{F}^{q+n} \Omega^{(q|2)} \left( \mathfrak{g} \right) = 0 \ , \ \forall n \in \mathbb{N} \setminus \left\lbrace 0 \right\rbrace , q \in \mathbb{Z} \ , \\
	\label{TSSZC} \tilde{F}^{p+1} \Omega^{(q|2)} \left( \mathfrak{g} \right) &\subseteq \tilde{F}^{p} \Omega^{(q|2)} \left( \mathfrak{g} \right) \ , \ \forall p,q \in \mathbb{Z} \ ,
\end{align}
analogously to \eqref{TSSB} and \eqref{TSSC}.

Again, we should verify that \eqref{TSSZA} correctly defines a filtration, namely
\begin{equation}\label{TSSZD}
	d \tilde{F}^p \Omega^{(q|2)} \left( \mathfrak{g} \right) \subseteq \tilde{F}^p \Omega^{(q+1|2)} \left( \mathfrak{g} \right) \ , \ \forall p,q \in \mathbb{Z} \ .
\end{equation}
We can verify this for $p=q$, and by using the same manipulations the general proof follows. We have
\begin{equation}\label{TSSZE}
	\tilde{F}^q \Omega^{(q|2)} \left( \mathfrak{g} \right) = \left\lbrace \omega \in \Omega^{(q|2)} \left( \mathfrak{g} \right) : \forall \xi^*_i \in  \mathfrak{h}^* , 
	\xi^*_{i} \wedge \omega = 0 \right\rbrace \ , $$ $$ \tilde{F}^q \Omega^{(q+1|2)} \left( \mathfrak{g} \right) = \left\lbrace \omega \in \Omega^{(q+1|2)} \left( \mathfrak{g} \right) : \forall \xi^*_i \in  \mathfrak{h}^* , \xi^*_{i_1} \wedge \xi^*_{i_{2}} \wedge \omega = 0 \right\rbrace \ .
\end{equation}
In order to verify that $d \omega \in \tilde{F}^q \Omega^{(q+1|2)} \left( \mathfrak{g} \right)$, we have to check that $\xi^*_{i_1} \wedge \xi^*_{i_{2}} \wedge d\omega = 0, \forall \xi^*_{i_1} , \xi^*_{i_{2}} \in \mathfrak{h}^*$. In particular, we have
\begin{equation}\label{TSSZF}
	\xi^*_{i_1} \wedge \xi^*_{i_{2}} \wedge d\omega = \xi^*_{i_1} \wedge \left[ \left( -1 \right)^{\left| \xi^*_{i_2} \right|} d \left( \xi^*_{i_2} \wedge \omega \right) - \left( -1 \right)^{\left| \xi^*_{i_2} \right|} \left( d \xi^*_{i_2} \right) \wedge \omega \right] = \left( -1 \right)^{\left| \xi^*_{i_1} \right| + \left| \xi^*_{i_2} \right|} \left( d \xi^*_{i_2} \right) \wedge \xi^*_{i_1} \wedge \omega = 0 \ ,
\end{equation}
where we used the Leibniz rule $\displaystyle d \left( \xi^*_{i_2} \wedge \omega \right) = \left( d \xi^*_{i_2} \right) \wedge \omega + \left( -1 \right)^{\left| \xi^*_{i_2} \right|} \xi^*_{i_2} \wedge \left( d \omega \right)$ and the fact that $\omega \in \tilde{F}^q \Omega^{(q|2)} \left( \mathfrak{g} \right)$. \eqref{TSSZA} correctly defines a filtration on $\Omega^{(\bullet|2)} \left( \mathfrak{g} \right)$.

In table \eqref{TableTSSD} we write explicitly the spaces of the filtration defined in \eqref{TSSZA}. Analogously to \eqref{TSSH}, we use the notation
\begin{equation}\label{TSSZG}
	\tilde{F}^{p} \Omega^{(q|2)} \left( \mathfrak{g} \right) \equiv \Omega^{(q|2)}_{* 2q-p-3} \left( \mathfrak{g} \right) \defeq \bigoplus_{i=0}^{q-p} \Omega^{(q-3+i|0)} \left( \mathfrak{k} \right) \otimes \Omega^{(3-i|2)} \left( \mathfrak{h} \right) \ ,
	%\Omega^{(q|2)}_{* i} \left( \mathfrak{g} \right) = \tilde{F}^{q-i} \Omega^{(q|2)} \left( \mathfrak{g} \right) 
	%\defeq \left\lbrace \omega \in \Omega^{(p|2)} \left( \mathfrak{g} \right) : \forall \xi^*_l \in  \mathfrak{h}^*, \xi^*_{1} \wedge \ldots \wedge \xi^*_{i+1} \wedge \omega = 0 \right\rbrace \ ,
\end{equation}
that is, $\displaystyle \Omega^{(q|2)}_{* 2q-p-3} \left( \mathfrak{g} \right) $ is the space of $\displaystyle \Omega^{(q|2)} \left( \mathfrak{g} \right)$ forms with \emph{at most} $2q-p-3$ superforms of 
$\mathfrak{k}^*$. We report \eqref{TSSZG} in Table \ref{TableTSSD}. Notice that $\Omega^{(3-i|2)} \left( \mathfrak{h} \right)$ is obtained by acting with $i$ contractions on the Berezin module of the sub-superalgebra $\mathfrak{h}$ (actually, one has $\mathpzc{B}er \left( \mathfrak{h} \right) \otimes Sym^{i} \Pi \mathfrak{h}$); in this way we can conveniently interpret $\Omega^{(q|2)}_{\ast 2q-p-3} \left( \mathfrak{g} \right)$ in \eqref{TSSZG} as a space which depends \emph{at most} on $q-p$ contractions along vectors in $\mathfrak{h}$.
\begin{table}[ht!]
\centering
\begin{tabular}{cccccccccc}
\multicolumn{1}{c|}{$\ldots$} & \multicolumn{1}{c|}{$\ldots$} & \multicolumn{1}{c|}{$\ldots$}                                      & \multicolumn{1}{c|}{$\vdots$} & \multicolumn{1}{c|}{$\ldots$}                                                      & \multicolumn{1}{c|}{$\ldots$}                                                      & \multicolumn{1}{c|}{$\ldots$}                                                      & \multicolumn{1}{c|}{$\ldots$}                                                      & \multicolumn{1}{c|}{$\ldots$}                                                                                                         & $\ldots$ \\ \cline{1-3} \cline{5-10} 
\multicolumn{1}{c|}{$\ldots$} & \multicolumn{1}{c|}{0}        & \multicolumn{1}{c|}{0}                                             & \multicolumn{1}{c|}{5}        & \multicolumn{1}{c|}{0}                                                             & \multicolumn{1}{c|}{0}                                                             & \multicolumn{1}{c|}{0}                                                             & \multicolumn{1}{c|}{0}                                                             & \multicolumn{1}{c|}{0}                                                                                                                & $\ldots$ \\ \cline{1-3} \cline{5-10} 
\multicolumn{1}{c|}{$\ldots$} & \multicolumn{1}{c|}{0}        & \multicolumn{1}{c|}{0}                                             & \multicolumn{1}{c|}{4}        & \multicolumn{1}{c|}{0}                                                             & \multicolumn{1}{c|}{0}                                                             & \multicolumn{1}{c|}{0}                                                             & \multicolumn{1}{c|}{0}                                                             & \multicolumn{1}{c|}{$\Omega^{(3|2)} \left( \mathfrak{h} \right) \otimes \Omega^{(1|0)} \left( \mathfrak{k} \right)$}                  & $\ldots$ \\ \cline{1-3} \cline{5-10} 
\multicolumn{1}{c|}{$\ldots$} & \multicolumn{1}{c|}{0}        & \multicolumn{1}{c|}{0}                                             & \multicolumn{1}{c|}{3}        & \multicolumn{1}{c|}{0}                                                             & \multicolumn{1}{c|}{0}                                                             & \multicolumn{1}{c|}{0}                                                             & \multicolumn{1}{c|}{$\Omega^{(3|2)} \left( \mathfrak{h} \right)$}                  & \multicolumn{1}{c|}{$\Omega^{(3|2)}_{ 1} \left( \mathfrak{g} \right) \otimes \Omega^{(1|0)} \left( \mathfrak{k} \right)$} & $\ldots$ \\ \cline{1-3} \cline{5-10} 
\multicolumn{1}{c|}{$\ldots$} & \multicolumn{1}{c|}{0}        & \multicolumn{1}{c|}{0}                                             & \multicolumn{1}{c|}{2}        & \multicolumn{1}{c|}{0}                                                             & \multicolumn{1}{c|}{0}                                                             & \multicolumn{1}{c|}{0}                                                             & \multicolumn{1}{c|}{$\Omega^{(3|2)}_{ 1} \left( \mathfrak{g} \right)$} & \multicolumn{1}{c|}{$\Omega^{(3|2)}_{ 2} \left( \mathfrak{g} \right) \otimes \Omega^{(1|0)} \left( \mathfrak{k} \right)$} & $\ldots$ \\ \cline{1-3} \cline{5-10} 
\multicolumn{1}{c|}{$\ldots$} & \multicolumn{1}{c|}{0}        & \multicolumn{1}{c|}{0}                                             & \multicolumn{1}{c|}{1}        & \multicolumn{1}{c|}{0}                                                             & \multicolumn{1}{c|}{0}                                                             & \multicolumn{1}{c|}{$\Omega^{(2|2)} \left( \mathfrak{h} \right)$}                  & \multicolumn{1}{c|}{$\Omega^{(3|2)}_{ 2} \left( \mathfrak{g} \right)$} & \multicolumn{1}{c|}{$\Omega^{(3|2)}_{ 3} \left( \mathfrak{g} \right) \otimes \Omega^{(1|0)} \left( \mathfrak{k} \right)$} & $\ldots$ \\ \cline{1-3} \cline{5-10} 
\multicolumn{1}{c|}{$\ldots$} & \multicolumn{1}{c|}{0}        & \multicolumn{1}{c|}{0}                                             & \multicolumn{1}{c|}{0}        & \multicolumn{1}{c|}{0}                                                             & \multicolumn{1}{c|}{0}                                                             & \multicolumn{1}{c|}{$\Omega^{(2|2)}_{ 1} \left( \mathfrak{g} \right)$} & \multicolumn{1}{c|}{$\Omega^{(3|2)}_{ 3} \left( \mathfrak{g} \right)$} & \multicolumn{1}{c|}{$\Omega^{(3|2)}_{ 4} \left( \mathfrak{g} \right) \otimes \Omega^{(1|0)} \left( \mathfrak{k} \right)$} & $\ldots$ \\ \cline{1-3} \cline{5-10} 
$\ldots$                      & -2                            & -1                                                                 &                               & 0                                                                                  & 1                                                                                  & 2                                                                                  & 3                                                                                  & 4                                                                                                                                     & $\ldots$ \\ \cline{1-3} \cline{5-10} 
\multicolumn{1}{c|}{$\ldots$} & \multicolumn{1}{c|}{0}        & \multicolumn{1}{c|}{0}                                             & \multicolumn{1}{c|}{-1}       & \multicolumn{1}{c|}{0}                                                             & \multicolumn{1}{c|}{$\Omega^{(1|2)} \left( \mathfrak{h} \right)$}                  & \multicolumn{1}{c|}{$\Omega^{(2|2)}_{ 2} \left( \mathfrak{g} \right)$} & \multicolumn{1}{c|}{$\Omega^{(3|2)}_{ 4} \left( \mathfrak{g} \right)$} & \multicolumn{1}{c|}{$\Omega^{(3|2)}_{ 5} \left( \mathfrak{g} \right) \otimes \Omega^{(1|0)} \left( \mathfrak{k} \right)$} & $\ldots$ \\ \cline{1-3} \cline{5-10} 
\multicolumn{1}{c|}{$\ldots$} & \multicolumn{1}{c|}{0}        & \multicolumn{1}{c|}{0}                                             & \multicolumn{1}{c|}{-2}       & \multicolumn{1}{c|}{0}                                                             & \multicolumn{1}{c|}{$\Omega^{(1|2)}_{ 1} \left( \mathfrak{g} \right)$} & \multicolumn{1}{c|}{$\Omega^{(2|2)}_{ 3} \left( \mathfrak{g} \right)$} & \multicolumn{1}{c|}{$\Omega^{(3|2)}_{ 5} \left( \mathfrak{g} \right)$} & \multicolumn{1}{c|}{$\Omega^{(3|2)}_{ 6} \left( \mathfrak{g} \right) \otimes \Omega^{(1|0)} \left( \mathfrak{k} \right)$} & $\ldots$ \\ \cline{1-3} \cline{5-10} 
\multicolumn{1}{c|}{$\ldots$} & \multicolumn{1}{c|}{0}        & \multicolumn{1}{c|}{0}                                             & \multicolumn{1}{c|}{-3}       & \multicolumn{1}{c|}{$\Omega^{(0|2)} \left( \mathfrak{h} \right)$}                  & \multicolumn{1}{c|}{$\Omega^{(1|2)}_{ 2} \left( \mathfrak{g} \right)$} & \multicolumn{1}{c|}{$\Omega^{(2|2)}_{ 4} \left( \mathfrak{g} \right)$} & \multicolumn{1}{c|}{$\Omega^{(3|2)}_{ 6} \left( \mathfrak{g} \right)$} & \multicolumn{1}{c|}{$\Omega^{(3|2)}_{ 7} \left( \mathfrak{g} \right) \otimes \Omega^{(1|0)} \left( \mathfrak{k} \right)$} & $\ldots$ \\ \cline{1-3} \cline{5-10} 
\multicolumn{1}{c|}{$\ldots$} & \multicolumn{1}{c|}{0}        & \multicolumn{1}{c|}{0}                                             & \multicolumn{1}{c|}{-4}       & \multicolumn{1}{c|}{$\Omega^{(0|2)}_{ 1} \left( \mathfrak{g} \right)$} & \multicolumn{1}{c|}{$\Omega^{(1|2)}_{ 3} \left( \mathfrak{g} \right)$} & \multicolumn{1}{c|}{$\Omega^{(2|2)}_{ 5} \left( \mathfrak{g} \right)$} & \multicolumn{1}{c|}{$\Omega^{(3|2)}_{ 7} \left( \mathfrak{g} \right)$} & \multicolumn{1}{c|}{$\Omega^{(3|2)}_{ 8} \left( \mathfrak{g} \right) \otimes \Omega^{(1|0)} \left( \mathfrak{k} \right)$} & $\ldots$ \\ \cline{1-3} \cline{5-10} 
\multicolumn{1}{c|}{$\ldots$} & \multicolumn{1}{c|}{0}        & \multicolumn{1}{c|}{$\Omega^{(-1|2)} \left( \mathfrak{h} \right)$} & \multicolumn{1}{c|}{-5}       & \multicolumn{1}{c|}{$\Omega^{(0|2)}_{ 2} \left( \mathfrak{g} \right)$} & \multicolumn{1}{c|}{$\Omega^{(1|2)}_{ 4} \left( \mathfrak{g} \right)$} & \multicolumn{1}{c|}{$\Omega^{(2|2)}_{ 6} \left( \mathfrak{g} \right)$} & \multicolumn{1}{c|}{$\Omega^{(3|2)}_{ 8} \left( \mathfrak{g} \right)$} & \multicolumn{1}{c|}{$\Omega^{(3|2)}_{ 9} \left( \mathfrak{g} \right) \otimes \Omega^{(1|0)} \left( \mathfrak{k} \right)$} & $\ldots$ \\ \cline{1-3} \cline{5-10} 
\multicolumn{1}{c|}{$\ldots$} & \multicolumn{1}{c|}{$\ldots$} & \multicolumn{1}{c|}{$\ldots$}                                      & \multicolumn{1}{c|}{$\vdots$} & \multicolumn{1}{c|}{$\ldots$}                                                      & \multicolumn{1}{c|}{$\ldots$}                                                      & \multicolumn{1}{c|}{$\ldots$}                                                      & \multicolumn{1}{c|}{$\ldots$}                                                      & \multicolumn{1}{c|}{$\ldots$}                                                                                                         & $\ldots$
\end{tabular}
\caption{Filtration defined in \eqref{TSSZA}. The integers $q$ and $p$ are spanned on the horizontal and vertical axes, respectively.}\label{TableTSSD}
\end{table}

Notice, once again, that the construction of table \ref{TableTSSD} is easily generalisable. We used the fact that $\mathfrak{g} = \mathfrak{osp}(2|2)$ and that $\mathfrak{h}=\mathfrak{osp}(1|2)$ to set the upper bound of $\Omega^{(q|2)} \left( \mathfrak{h} \right)$ to $\Omega^{(3|2)} \left( \mathfrak{h} \right)$ for $q \geq 3$. The construction of the filtration will be generalised in the following section.

From \eqref{TSSZA} we can easily construct page 0 of the spectral sequence. It is more convenient to shift  the spaces in order to display page 0 in an analogous fashion to that of Table \ref{TableTSSB}:
\begin{equation}\label{TSSZI}
	\begin{cases}
		m \to m - 3 +2n \ , \\
		n \to 3-n \ ,
	\end{cases} \ \implies \ \tilde{E}_0^{m,n} \to \tilde{E}_0^{m,n} = \tilde{F}^{m-3+2n} \Omega^{(m+n|2)} \left( \mathfrak{g} \right) / \tilde{F}^{m-2+2n} \Omega^{(m+n|2)} \left( \mathfrak{g} \right) \ .
\end{equation}
Page 0 then reads as in table \ref{TableTSSF}.
\begin{table}[ht!]
\centering 
\begin{tabular}{ccccccc}                            
& \multicolumn{1}{c}0                                                                                                                    
& \multicolumn{1}{c}1                                                                                                                    
& \multicolumn{1}{c}2                                                                                                                    
& \multicolumn{1}{c}3                                                                                                                    
& \multicolumn{1}{c}4                             
& \multicolumn{1}{c}{$\ldots$} 
\\ \cline{2-7} 
 \multicolumn{1}{c|}{$\vdots$} 
& \multicolumn{1}{c|}{$\ldots$}                                                                                        
& \multicolumn{1}{c|}{$\ldots$}                                                                                        
& \multicolumn{1}{c|}{$\ldots$}                                                                                        
& \multicolumn{1}{c|}{$\ldots$}                                                                                       
& \multicolumn{1}{c|}{$\ldots$} 
& $\ldots$ 
\\  \cline{2-7} 
 \multicolumn{1}{c|}{4}        
& \multicolumn{1}{c|}{$\Omega^{(0|2)} \left( \mathfrak{h} \right) \otimes \Omega^{(4|0)} \left( \mathfrak{k} \right)$} 
& \multicolumn{1}{c|}{$\Omega^{(1|2)} \left( \mathfrak{h} \right) \otimes \Omega^{(4|0)} \left( \mathfrak{k} \right)$} 
& \multicolumn{1}{c|}{$\Omega^{(2|2)} \left( \mathfrak{h} \right) \otimes \Omega^{(4|0)} \left( \mathfrak{k} \right)$} 
& \multicolumn{1}{c|}{$\Omega^{(3|2)} \left( \mathfrak{h} \right) \otimes \Omega^{(4|0)} \left( \mathfrak{k} \right)$} 
& \multicolumn{1}{c|}{0}        
& $\ldots$ 
\\ \cline{2-7}  
 \multicolumn{1}{c|}{3}        
& \multicolumn{1}{c|}{$\Omega^{(0|2)} \left( \mathfrak{h} \right) \otimes \Omega^{(3|0)} \left( \mathfrak{k} \right)$} 
& \multicolumn{1}{c|}{$\Omega^{(1|2)} \left( \mathfrak{h} \right) \otimes \Omega^{(3|0)} \left( \mathfrak{k} \right)$} 
& \multicolumn{1}{c|}{$\Omega^{(2|2)} \left( \mathfrak{h} \right) \otimes \Omega^{(3|0)} \left( \mathfrak{k} \right)$} 
& \multicolumn{1}{c|}{$\Omega^{(3|2)} \left( \mathfrak{h} \right) \otimes \Omega^{(3|0)} \left( \mathfrak{k} \right)$} 
& \multicolumn{1}{c|}{0}        
& $\ldots$ 
\\ \cline{2-7}
 \multicolumn{1}{c|}{2}        
& \multicolumn{1}{c|}{$\Omega^{(0|2)} \left( \mathfrak{h} \right) \otimes \Omega^{(2|0)} \left( \mathfrak{k} \right)$} 
& \multicolumn{1}{c|}{$\Omega^{(1|2)} \left( \mathfrak{h} \right) \otimes \Omega^{(2|0)} \left( \mathfrak{k} \right)$} 
& \multicolumn{1}{c|}{$\Omega^{(2|2)} \left( \mathfrak{h} \right) \otimes \Omega^{(2|0)} \left( \mathfrak{k} \right)$} 
& \multicolumn{1}{c|}{$\Omega^{(3|2)} \left( \mathfrak{h} \right) \otimes \Omega^{(2|0)} \left( \mathfrak{k} \right)$} 
& \multicolumn{1}{c|}{0}        
& $\ldots$ 
\\ \cline{2-7} 
\multicolumn{1}{c|}{1}        
& \multicolumn{1}{c|}{$\Omega^{(0|2)} \left( \mathfrak{h} \right) \otimes \Omega^{(1|0)} \left( \mathfrak{k} \right)$} 
& \multicolumn{1}{c|}{$\Omega^{(1|2)} \left( \mathfrak{h} \right) \otimes \Omega^{(1|0)} \left( \mathfrak{k} \right)$} 
& \multicolumn{1}{c|}{$\Omega^{(2|2)} \left( \mathfrak{h} \right) \otimes \Omega^{(1|0)} \left( \mathfrak{k} \right)$} 
& \multicolumn{1}{c|}{$\Omega^{(3|2)} \left( \mathfrak{h} \right) \otimes \Omega^{(1|0)} \left( \mathfrak{k} \right)$} 
& \multicolumn{1}{c|}{0}        
& $\ldots$ 
\\ \cline{2-7} 
 \multicolumn{1}{c|}{0}        
& \multicolumn{1}{c|}{$\Omega^{(0|2)} \left( \mathfrak{h} \right)$}                                                    
& \multicolumn{1}{c|}{$\Omega^{(1|2)} \left( \mathfrak{h} \right)$}                                                    
& \multicolumn{1}{c|}{$\Omega^{(2|2)} \left( \mathfrak{h} \right)$}                                                    
& \multicolumn{1}{c|}{$\Omega^{(3|2)} \left( \mathfrak{h} \right)$}                                                    
& \multicolumn{1}{c|}{0}        
& $\ldots$ 
\\ \cline{2-7}  
 \multicolumn{1}{c|}{-1}       
& \multicolumn{1}{c|}{0}                                                                                               
& \multicolumn{1}{c|}{0}                                                                                               
& \multicolumn{1}{c|}{0}                                                                                               
& \multicolumn{1}{c|}{0}                                                                                               
& \multicolumn{1}{c|}{0}        
& $\ldots$ 
\\ \cline{2-7} 
 \multicolumn{1}{c|}{$\vdots$} 
& \multicolumn{1}{c|}{$\ldots$}                                                                                        
& \multicolumn{1}{c|}{$\ldots$}                                                                                        
& \multicolumn{1}{c|}{$\ldots$}                                                                                        
& \multicolumn{1}{c|}{$\ldots$}                                                                                        
& \multicolumn{1}{c|}{$\ldots$} 
& $\ldots$
\end{tabular}
\caption{$\tilde{E}_0^{m,n}$ as defined in \eqref{TSSZI}. The integers $n$ and $m$ are spanned on the horizontal and vertical axes, respectively.}\label{TableTSSF}
\end{table}

Notice that, by contrast with table \ref{TableTSSB}, table \ref{TableTSSF} has non-trivial entries in the first and second quadrants, as a consequence of the facts that integral forms of $\mathfrak{h}$ are unbounded from below and that superforms of $\mathfrak{k}$ are bounded from below. The differentials that we use to define the next pages of the spectral sequence are the same as before, therefore we can construct page 1 as
\begin{equation}\label{TSSZJ}
	d_0 = V_{\mathfrak{k}} V_{\mathfrak{h}} \iota_{\mathfrak{k}} + V_{\mathfrak{h}} V_{\mathfrak{h}} \iota_{\mathfrak{h}} \ , \ \implies \ \tilde{E}_1^{m,n} \defeq H \left( \tilde{E}_0^{m,n} , d_0 \right) \ .
\end{equation}
In exact analogy to the previous case, we have
\begin{equation}\label{TSSZK}
	H \left( \Omega^{(p|2)} \left( \mathfrak{h} \right) \otimes \Omega^{(q|0)} \left( \mathfrak{k} \right) , d_0 \right) = H^{(p|2)} \left( \mathfrak{h} \right) \otimes \left( \Omega^{(q|0)} \left( \mathfrak{k} \right) \right)^{\mathfrak{h}} \ .
\end{equation}
In the specific case under examination, we have
\begin{equation}\label{TSSZL}
	H^{(p|2)} \left( \mathfrak{h} \right) = \begin{cases}
		\mathbb{R} \ , \ \text{if} \ p = 0 \ ,\\
		\Pi \mathbb{R} \ , \ \text{if} \ p = 3 \ ,\\
		\left\lbrace 0 \right\rbrace \ , \ \text{else} \ ,
	\end{cases} \ , \ \left( \Omega^{(q|0)} \left( \mathfrak{k} \right) \right)^{\mathfrak{h}} = \begin{cases}
		\mathbb{R} \ , \ \text{if} \ q = 0 \ ,\\
		\left\lbrace 0 \right\rbrace \ , \ \text{else} \ .
	\end{cases}
\end{equation}
In particular, the cohomology spaces $H^{(p|2)} \left( \mathfrak{h} \right)$ can be easily determined starting from the spaces $H^{(p|0)} \left( \mathfrak{h} \right)$, via the map \eqref{VSPQ}:
\begin{equation}\label{TSSZLA}
	H^{(p|0)} \left( \mathfrak{h} \right) = \begin{cases}
		\mathbb{R} \ , \ \text{if} \ p = 0 \ ,\\
		\Pi \mathbb{R} \ , \ \text{if} \ p = 3 \ ,\\
		\left\lbrace 0 \right\rbrace \ , \ \text{else} \ ,
	\end{cases} \ \overset{\star}{\underset{}{\longleftrightarrow}} \ \ H^{(3-p|2)} \left( \mathfrak{h} \right) = \begin{cases}
		\Pi \mathbb{R} \ , \ \text{if} \ p = 0 \ ,\\
		\mathbb{R} \ , \ \text{if} \ p = 3 \ ,\\
		\left\lbrace 0 \right\rbrace \ , \ \text{else} \ ,
	\end{cases} \ .
\end{equation}

As we discussed above, this already gives page 2 of the spectral sequence, since the differential $d_1$ is trivial:
\begin{equation}\label{TSSZLAA}
	\tilde{E}^{m,n}_2 \defeq H \left( \tilde{E}_1^{m,n} , d_1 \right) = \tilde{E}_1^{m,n} \ .
\end{equation}
In Table \ref{TableTSSG} we write page $\tilde{E}_2^{m,n}$:
\begin{table}[ht!]
\centering
\begin{tabular}{cccccccc}
\multicolumn{1}{c|}{$\ldots$} & \multicolumn{1}{c|}{$\vdots$} & \multicolumn{1}{c|}{$\ldots$}                                & \multicolumn{1}{c|}{$\ldots$} & \multicolumn{1}{c|}{$\ldots$} & \multicolumn{1}{c|}{$\ldots$}                                & \multicolumn{1}{c|}{$\ldots$} & $\ldots$ \\ \cline{1-1} \cline{3-8} 
\multicolumn{1}{c|}{$\ldots$} & \multicolumn{1}{c|}{2}        & \multicolumn{1}{c|}{0}                                       & \multicolumn{1}{c|}{0}        & \multicolumn{1}{c|}{0}        & \multicolumn{1}{c|}{0}                                       & \multicolumn{1}{c|}{0}        & $\ldots$ \\ \cline{1-1} \cline{3-8} 
\multicolumn{1}{c|}{$\ldots$} & \multicolumn{1}{c|}{1}        & \multicolumn{1}{c|}{0}                                       & \multicolumn{1}{c|}{0}        & \multicolumn{1}{c|}{0}        & \multicolumn{1}{c|}{0}                                       & \multicolumn{1}{c|}{0}        & $\ldots$ \\ \cline{1-1} \cline{3-8} 
\multicolumn{1}{c|}{$\ldots$} & \multicolumn{1}{c|}{0}        & \multicolumn{1}{c|}{$H^{(0|2)} \left( \mathfrak{h} \right)$} & \multicolumn{1}{c|}{0}        & \multicolumn{1}{c|}{0}        & \multicolumn{1}{c|}{$H^{(3|2)} \left( \mathfrak{h} \right)$} & \multicolumn{1}{c|}{0}        & $\ldots$ \\ \cline{1-1} \cline{3-8} 
$\ldots$                      &                               & 0                                                            & 1                             & 2                             & 3                                                            & 4                             & $\ldots$ \\ \cline{1-1} \cline{3-8} 
\multicolumn{1}{c|}{$\ldots$} & \multicolumn{1}{c|}{-1}       & \multicolumn{1}{c|}{0}                                       & \multicolumn{1}{c|}{0}        & \multicolumn{1}{c|}{0}        & \multicolumn{1}{c|}{0}                                       & \multicolumn{1}{c|}{0}        & $\ldots$ \\ \cline{1-1} \cline{3-8} 
\multicolumn{1}{c|}{$\ldots$} & \multicolumn{1}{c|}{$\vdots$} & \multicolumn{1}{c|}{$\ldots$}                                & \multicolumn{1}{c|}{$\ldots$} & \multicolumn{1}{c|}{$\ldots$} & \multicolumn{1}{c|}{$\ldots$}                                & \multicolumn{1}{c|}{$\ldots$} & $\ldots$
\end{tabular}
\caption{$\tilde{E}_2^{m,n}$ as defined in \eqref{TSSZLAA}. The integers $n$ and $m$ are spanned on the horizontal and vertical axes, respectively.}\label{TableTSSG}
\end{table}

The other differentials all vanish as in the previous case, hence the spectral sequence converges at page $\tilde{E}_1^{m,n} = \tilde{E}_2^{m,n} = \ldots = \tilde{E}_\infty^{m,n}$, and in particular
\begin{equation}\label{TSSZM}
	H^{(0|2)} \left( \mathfrak{g} \right) = \mathbb{R} \ , \ H^{(3|2)} \left( \mathfrak{g} \right) = \Pi \mathbb{R} \ .
\end{equation}

We can read the result in two ways: by using the filtration introduced in \eqref{TSSZA}, we have confirmed that there are non-trivial cohomology classes among pseudoforms, 
but also shown that the classes we found are inequivalent to those found in \eqref{TSSY}. This is consistent with the inequivalence of Clifford-Weyl modules defined in \cite{Witten}, as mentioned before.

We can put all the results together to list the \emph{full} algebraic cohomology for the example $\mathfrak{osp} \left( 2|2 \right)$:
\begin{align}
	\label{TSSZN} H^p_{super} \left( \mathfrak{osp} \left( 2|2 \right) \right) &= \begin{cases}
		\Pi^p \mathbb{R} \ , \ \text{if} \ p=0,3 \ , \\
		\left\lbrace 0 \right\rbrace \ , \ \text{else} \ ,
	\end{cases} \\
	\label{TSSZO} H^p_{pseudo,2} \left( \mathfrak{osp} \left( 2|2 \right) \right) &= \begin{cases}
		\Pi^p \mathbb{R} \ , \ \text{if} \ p=0,1,3,4 \ , \\
		\left\lbrace 0 \right\rbrace \ , \ \text{else} \ ,
	\end{cases} \\
	\label{TSSZP} H^p_{integral} \left( \mathfrak{osp} \left( 2|2 \right) \right) &= \begin{cases}
		\Pi^p \mathbb{R} \ , \ \text{if} \ p=1,4 \ , \\
		\left\lbrace 0 \right\rbrace \ , \ \text{else} \ ,
	\end{cases}
\end{align}
where we indicated $\Pi^p = \Pi$ for $p=1 \mod 2$, $\Pi^p = \text{id}$ for $p=0 \mod 2$ and the subscript \virgolette pseudo,2" is used to emphasise that the pseudoforms are obtained at picture number 2.

%\begin{remark}
	From \eqref{TSSZN}, \eqref{TSSZO} and \eqref{TSSZP} we see that the duality between superforms and integral forms given by the \virgolette $\star$" map extends (at least for the given example) to the complex of pseudoforms as well. This is a consequence of the fact that we had to introduce two inequivalent filtrations and calculate the cohomology in the two cases. We conjecture that this result should hold for any basic Lie superalgebra. This would represent the extension to Lie superalgebras of the Poincar\'e duality: in Lie algebras, one has
	\begin{equation}
		H^p \left( \mathfrak{g} \right) \cong H^{n-p} \left( \mathfrak{g} \right) \ ,
	\end{equation}
where the top form $\omega \in H^{n} \left( \mathfrak{g} \right)$ is the fulcrum of the duality. In \cite{CCGN2} the authors used the same argument to demonstrate the isomorphism between superform and integral form cohomologies. This shows that, if Poincar\'e duality can be extended to superalgebras, it may involve forms in different complexes, i.e., with different picture number. We then conjecture the existence of an isomorphism
	\begin{equation}
		\star : H^{(\bullet|\bullet)} \left( \mathfrak{g} , \mathbb{R} \right) \overset{\cong}{\underset{}{\longrightarrow}} H^{(m-\bullet|n-\bullet)} \left( \mathfrak{g} , \mathbb{R} \right) \ ,
	\end{equation}
	which would extend the Poincar\'e duality to superalgebras. In the following section we will introduce the definition of a generalised page zero that implements both the pseudoforms emerging from $\mathfrak{k}$ and the pseudoforms emerging from $\mathfrak{h}$.
%\end{remark}

\section{General Constructions}

In the previous sections we have shown how to introduce two inequivalent filtrations that we have used to calculate the algebraic cohomology, in particular the pseudoform cohomology, for the specific example $\mathfrak{osp}(2|2)$. We have shown that \emph{both} filtrations should be considered in order to reproduce all the possible pseudoforms. As we emphasised many times, the construction of such filtrations is very general and can be extended to any superalgebra $\mathfrak{g}$ with a sub-superalgebra $\mathfrak{h}$ and for any picture number $l$. Here we deal with the general constructions.

We start by generalising the definitions \eqref{TSSA} and \eqref{TSSZA} that keep into account both the sectors that define page 0 of the spectral sequence: 
\begin{eqnarray}
	\label{GCA} F^p \Omega^{(q|l)} \left( \mathfrak{g} \right) &\defeq& \left\lbrace \omega \in \Omega^{(q|l)} \left( \mathfrak{g} \right) : \forall \xi_i \in \mathfrak{h} , \iota_{\xi_{i_1}} \ldots \iota_{\xi_{i_{q+1-p}}} \omega = 0 \right\rbrace \ , \\
	\label{GCAA} \tilde{F}^{p} \Omega^{(q|l)} \left( \mathfrak{g} \right) &\defeq& \left\lbrace \omega \in \Omega^{(q|2)} \left( \mathfrak{g} \right) : \forall \xi^*_i \in \mathfrak{h}^* , 
	\xi^*_{i_1} \wedge \ldots \wedge \xi^*_{i_{q+1-p}} \wedge \omega = 0 \right\rbrace \ ,
\end{eqnarray}
where $l$ is a generic picture number.
In this way we can define the \emph{total} page zero of the spectral sequence, for each $l$, as
\begin{equation}\label{GCB}
	\mathcal{E}_0^{m,n} \defeq E_0^{m,n} \oplus \tilde{E}_0^{m,n} \defeq \frac{F^m \Omega^{(m+n|l)} \left( \mathfrak{g} \right)}{F^{m+1} \Omega^{(m+n|l)} \left( \mathfrak{g} \right)} \oplus \frac{\tilde{F}^{m+2n-r} \Omega^{(m+n|l)} \left( \mathfrak{g} \right)}{\tilde{F}^{m+2n-r+1} \Omega^{(m+n|l)} \left( \mathfrak{g} \right)} \ .
\end{equation}

%\begin{remark}
	Notice that the procedure actually refers also to the extremal cases of superforms and integral forms. In particular, we should observe that the definition \eqref{GCB} simplifies for $l=0$ and $\text{dim}_1 \left( \mathfrak{g} \right)$, the dimension of the odd subspace. In these cases, indeed, one directly sees that either the filtration $F^p\Omega^{(q|l)} \left( \mathfrak{g} \right)$ or the filtration $\tilde{F}^p\Omega^{(q|l)} \left( \mathfrak{g} \right)$ is empty (we are assuming that the sub-superalgebra has non-zero odd dimension):
	\begin{equation}\label{GCC}
		l = 0 \ \implies \ \tilde{F}^p \Omega^{(q|0)} \left( \mathfrak{g} \right) = \left\lbrace \omega \in \Omega^{(q|0)} \left( \mathfrak{g} \right) : \forall \xi^i \in \Pi \mathfrak{h}^* , \xi^{i_1} \wedge \ldots \wedge \xi^{i_{q+1-p}} \wedge \omega = 0 \right\rbrace = \left\lbrace 0 \right\rbrace \ ,
	\end{equation}
	as a consequence of the fact that there is no top superform;
	\begin{equation}\label{GCD}
		l = \text{dim}_1 \left( \mathfrak{g} \right) \ \implies \ F^p \Omega^{(q|\text{dim}_1 \left( \mathfrak{g} \right))} \left( \mathfrak{g} \right) = \left\lbrace \omega \in \Omega^{(q|\text{dim}_1 \left( \mathfrak{g} \right))} \left( \mathfrak{g} \right) : \forall \xi_i \in \mathfrak{h} , \iota_{\xi_{i_1}} \ldots \iota_{\xi_{i_{q+1-p}}} \omega = 0 \right\rbrace = \left\lbrace 0 \right\rbrace \ ,
	\end{equation}
	as a consequence of the fact that there is no bottom integral form. In these two cases, page zero in \eqref{GCB} then simplifies as:
	\begin{align}
		\label{GCE} l = 0 \ \implies \ & \mathcal{E}_0^{m,n} = E_0^{m,n} = \frac{F^m \Omega^{(m+n|l)} \left( \mathfrak{g} \right)}{F^{m+1} \Omega^{(m+n|0)} \left( \mathfrak{g} \right)} \ , \\
		\label{GCF} l = \text{dim}_1 \left( \mathfrak{g} \right) \ \implies \ & \mathcal{E}_0^{m,n} = \tilde{E}_0^{m,n} = \frac{\tilde{F}^{m+2n-r} \Omega^{(m+n|\text{dim}_1 \left( \mathfrak{g} \right))} \left( \mathfrak{g} \right)}{\tilde{F}^{m+2n-r+1} \Omega^{(m+n|\text{dim}_1 \left( \mathfrak{g} \right))} \left( \mathfrak{g} \right)} \ .
	\end{align}
%\end{remark}

%\begin{remark}
%	In \eqref{TSSZA} and in the paragraphs above, we commented on the necessity of introducing a complementary filtration (with respect to the usual one) by recalling \cite{Witten} and the fact that the modules introduced by contraction operators and multiplication by forms are inequivalent. Actually, this is true as long as one deals with \virgolette genuinely super" objects, i.e., in our case, with sub-superalgebras with non-trivial odd dimension. On the other hand, if one introduces a purely even sub-algebra $\mathfrak{h}$, the filtrations $F^p \Omega^{(q|l)} \left( \mathfrak{g} \right)$ and $\tilde{F}^p \Omega^{(q|l)} \left( \mathfrak{g} \right)$ induce
In \eqref{TSSZA} and in the paragraphs above, we emphasised that the complementary filtration related to the modules obtained by contraction operators and multiplication by forms are inequivalent. In particular, this is true for sub-superalgebras with non-trivial odd dimensions. On the other hand, if one introduces a purely even sub-algebra $\mathfrak{h}$, the filtrations $F^p \Omega^{(q|l)} \left( \mathfrak{g} \right)$ and $\tilde{F}^p \Omega^{(q|l)} \left( \mathfrak{g} \right)$ induce
	\begin{equation}\label{GCG}
		E_0^{m,n} \cong \tilde{E}_0^{m,n} \ ,
	\end{equation}
	 and $\mathcal{E}^{m,n}_0$ counts the spaces twice. The equivalence follows directly from the definitions \eqref{GCA} and \eqref{GCAA}, for any picture number $l$:
	 \begin{eqnarray}
	 	\label{GCGA} F^m \Omega^{(m+n|l)} \left( \mathfrak{g} \right) = \bigoplus_{i=0}^n \Omega^{(i|0)} \left( \mathfrak{h} \right) \otimes \Omega^{(m+n-i|l)} \left( \mathfrak{k} \right) \ &\implies& \ E_0^{m,n} = \Omega^{(n|0)} \left( \mathfrak{h} \right) \otimes \Omega^{(m|l)} \left( \mathfrak{k} \right) \ , \\
	 	\label{GCGB} \tilde{F}^{m+2n-r} \Omega^{(m+n|l)} \left( \mathfrak{g} \right) = \bigoplus_{i=n}^{\text{dim} \left( \mathfrak{h} \right)} \Omega^{(i|0)} \left( \mathfrak{h} \right) \otimes \Omega^{(m+n-i|l)} \left( \mathfrak{k} \right) \ &\implies& \ \tilde{E}_0^{m,n} = \Omega^{(n|0)} \left( \mathfrak{h} \right) \otimes \Omega^{(m|l)} \left( \mathfrak{k} \right) \ .
	 \end{eqnarray}
	
	It is interesting to note that this exactly corresponds to what is done for the the calculation of the algebraic superform cohomology for Lie superalgebras as in \cite{Fuks}. There, 
	one actually does not need to introduce the second filtration. The drawback is that one has to calculate pseudoform cohomology spaces explicitly, i.e., by using an explicit realisation. In the following section we show how to use the distributional realisation for pseudoforms and integral forms to calculate the cohomology spaces explicitly in the case of a bosonic sub-algebra.
	
	In \cite{Fuks}, the author uses $\mathfrak{h} = \mathfrak{g}_0$, and since the super-coset is purely odd, the commutation relations can be formally written as
	\begin{equation}\label{GCH}
		\left[ \mathfrak{h} , \mathfrak{h} \right] \subseteq \mathfrak{h} \ , \ \left[ \mathfrak{k} , \mathfrak{k} \right] \subseteq \mathfrak{h} \ , \ \left[ \mathfrak{h} , \mathfrak{k} \right] \subseteq \mathfrak{k} \ ,
	\end{equation}
	indicating that $\mathfrak{h}$ is reductive and $\mathfrak{k}$ is homogeneous. For the sake of clarity, we quickly review the superform cohomology in this case, for the example of the previous section $\mathfrak{osp}(2|2)$. We have that page zero is given by
	\begin{equation}\label{GCHA}
		E_0^{m,n} = \Omega^{(n|0)} \left( \mathfrak{h} \right) \otimes \Omega^{(m|0)} \left( \mathfrak{k} \right) \ .
	\end{equation}
	It is not difficult to prove that, because of the structure \eqref{GCH}, page 2 of the spectral sequence at picture number 0 is given by
	\begin{equation}\label{GCHB}
		E_2^{m,n} = H^{(n|0)} \left( \mathfrak{h} \right) \otimes H^{(m|0)} \left( \mathfrak{g} , \mathfrak{h} \right) \ .
	\end{equation}
	In particular, one has
	\begin{equation}\label{GCHC}
		H^{(n|0)} \left( \mathfrak{h} \right) = H^{(n|0)} \left( \mathfrak{so}(2) \oplus \mathfrak{sp}(2) \right) = \begin{cases}
			\mathbb{R} \ , \ \text{if} \ n = 0,4 \ , \\
			\Pi \mathbb{R} \ , \ \text{if} \ n = 1,3 \ , \\
			\left\lbrace 0 \right\rbrace \ , \ \text{else}.
		\end{cases}
	\end{equation}
	On the other hand, one has that the relative cohomology of the coset $\mathfrak{k}$ is infinitely generated as (in the following section we will give explicit expressions for the generators)
	\begin{equation}\label{GCHD}
		H^{(m|0)} \left( \mathfrak{g} , \mathfrak{h} \right) = \begin{cases}
			\mathbb{R} \ , \ \text{if} \ n = 0,2,4,\ldots \ , \\
			\left\lbrace 0 \right\rbrace \ , \ \text{if} \ \text{else} \ .
		\end{cases}
	\end{equation}
	We report \eqref{GCHB} in Table \eqref{TableGCsuper}.
	\begin{table}[]
\centering
\resizebox{1.07\textwidth}{!}{\hspace{-1.4cm}
\begin{tabular}{cccccccccc}
\multicolumn{1}{c|}{$\ldots$} & \multicolumn{1}{c|}{$\ldots$} & \multicolumn{1}{c|}{$\vdots$} & \multicolumn{1}{c|}{$\ldots$}                                                                              & \multicolumn{1}{c|}{$\ldots$}                                                                              & \multicolumn{1}{c|}{$\ldots$} & \multicolumn{1}{c|}{$\ldots$}                                                                              & \multicolumn{1}{c|}{$\ldots$}                                                                              & \multicolumn{1}{c|}{$\ldots$} & $\ldots$ \\ \cline{1-2} \cline{4-10} 
\multicolumn{1}{c|}{$\ldots$} & \multicolumn{1}{c|}{0}        & \multicolumn{1}{c|}{4}        & \multicolumn{1}{c|}{$H^{(0|0)} \left( \mathfrak{h} \right) \otimes H^{(4|0)} \left( \mathfrak{k} \right)$} & \multicolumn{1}{c|}{$H^{(1|0)} \left( \mathfrak{h} \right) \otimes H^{(4|0)} \left( \mathfrak{k} \right)$} & \multicolumn{1}{c|}{0}        & \multicolumn{1}{c|}{$H^{(3|0)} \left( \mathfrak{h} \right) \otimes H^{(4|0)} \left( \mathfrak{k} \right)$} & \multicolumn{1}{c|}{$H^{(4|0)} \left( \mathfrak{h} \right) \otimes H^{(4|0)} \left( \mathfrak{k} \right)$} & \multicolumn{1}{c|}{0}        & $\ldots$ \\ \cline{1-2} \cline{4-10} 
\multicolumn{1}{c|}{$\ldots$} & \multicolumn{1}{c|}{0}        & \multicolumn{1}{c|}{3}        & \multicolumn{1}{c|}{0}                                                                                     & \multicolumn{1}{c|}{0}                                                                                     & \multicolumn{1}{c|}{0}        & \multicolumn{1}{c|}{0}                                                                                     & \multicolumn{1}{c|}{0}                                                                                     & \multicolumn{1}{c|}{0}        & $\ldots$ \\ \cline{1-2} \cline{4-10} 
\multicolumn{1}{c|}{$\ldots$} & \multicolumn{1}{c|}{0}        & \multicolumn{1}{c|}{2}        & \multicolumn{1}{c|}{$H^{(0|0)} \left( \mathfrak{h} \right) \otimes H^{(2|0)} \left( \mathfrak{k} \right)$} & \multicolumn{1}{c|}{$H^{(1|0)} \left( \mathfrak{h} \right) \otimes H^{(2|0)} \left( \mathfrak{k} \right)$} & \multicolumn{1}{c|}{0}        & \multicolumn{1}{c|}{$H^{(3|0)} \left( \mathfrak{h} \right) \otimes H^{(2|0)} \left( \mathfrak{k} \right)$} & \multicolumn{1}{c|}{$H^{(4|0)} \left( \mathfrak{h} \right) \otimes H^{(2|0)} \left( \mathfrak{k} \right)$} & \multicolumn{1}{c|}{0}        & $\ldots$ \\ \cline{1-2} \cline{4-10} 
\multicolumn{1}{c|}{$\ldots$} & \multicolumn{1}{c|}{0}        & \multicolumn{1}{c|}{1}        & \multicolumn{1}{c|}{0}                                                                                     & \multicolumn{1}{c|}{0}                                                                                     & \multicolumn{1}{c|}{0}        & \multicolumn{1}{c|}{0}                                                                                     & \multicolumn{1}{c|}{0}                                                                                     & \multicolumn{1}{c|}{0}        & $\ldots$ \\ \cline{1-2} \cline{4-10} 
\multicolumn{1}{c|}{$\ldots$} & \multicolumn{1}{c|}{0}        & \multicolumn{1}{c|}{0}        & \multicolumn{1}{c|}{$H^{(0|0)} \left( \mathfrak{h} \right) \otimes H^{(0|0)} \left( \mathfrak{k} \right)$} & \multicolumn{1}{c|}{$H^{(1|0)} \left( \mathfrak{h} \right) \otimes H^{(0|0)} \left( \mathfrak{k} \right)$} & \multicolumn{1}{c|}{0}        & \multicolumn{1}{c|}{$H^{(3|0)} \left( \mathfrak{h} \right) \otimes H^{(0|0)} \left( \mathfrak{k} \right)$} & \multicolumn{1}{c|}{$H^{(4|0)} \left( \mathfrak{h} \right) \otimes H^{(0|0)} \left( \mathfrak{k} \right)$} & \multicolumn{1}{c|}{0}        & $\ldots$ \\ \cline{1-2} \cline{4-10} 
$\ldots$                      & -1                            &                               & 0                                                                                                          & 1                                                                                                          & 2                             & 3                                                                                                          & 4                                                                                                          & 5                             & $\ldots$ \\ \cline{1-2} \cline{4-10} 
\multicolumn{1}{c|}{$\ldots$} & \multicolumn{1}{c|}{0}        & \multicolumn{1}{c|}{-1}       & \multicolumn{1}{c|}{0}                                                                                     & \multicolumn{1}{c|}{0}                                                                                     & \multicolumn{1}{c|}{0}        & \multicolumn{1}{c|}{0}                                                                                     & \multicolumn{1}{c|}{0}                                                                                     & \multicolumn{1}{c|}{0}        & $\ldots$ \\ \cline{1-2} \cline{4-10} 
\multicolumn{1}{c|}{$\ldots$} & \multicolumn{1}{c|}{$\ldots$} & \multicolumn{1}{c|}{$\vdots$} & \multicolumn{1}{c|}{$\ldots$}                                                                              & \multicolumn{1}{c|}{$\ldots$}                                                                              & \multicolumn{1}{c|}{$\ldots$} & \multicolumn{1}{c|}{$\ldots$}                                                                              & \multicolumn{1}{c|}{$\ldots$}                                                                              & \multicolumn{1}{c|}{$\ldots$} & $\ldots$
\end{tabular}
}
\caption{$E_2^{m,n}$ as defined in \eqref{GCHB}. The integers $n$ and $m$ are spanned on the horizontal and vertical axes, respectively.}\label{TableGCsuper}
\end{table}
	The differential $d_2$ formally reads as
	\begin{equation}\label{GCHE}
		d_2 = V_{\mathfrak{k}} V_{\mathfrak{k}} \iota_\mathfrak{h} \ ,
	\end{equation}
	hence moving in \eqref{TableGCsuper} vertically by two and horizontally on the left by one. Page three of the spectral sequence is defined as
	\begin{equation}\label{GCHF}
		E_3^{m,n} \defeq H \left( E_2^{m,n} , d_2 \right) \ ,
	\end{equation}
	and it coincides with the convergence of the spectral sequence since all the higher differentials are trivial. It is now not difficult to verify that
	\begin{equation}
		H^{(p|0)} \left( \mathfrak{osp}(2|2) \right) = H^p_{super} \left( \mathfrak{osp}(2|2) \right) = H^p \left( \mathfrak{sp}(2) \right) = \begin{cases}
			\mathbb{R} \ , \ \text{if} \ p=0 \ , \\
			\Pi \mathbb{R} \ , \ \text{if} \ p=3 \ , \\
			\left\lbrace 0 \right\rbrace \ , \ \text{else},
		\end{cases}
	\end{equation}
	as demonstrated in \cite{Fuks}.
%\end{remark}

Let us now comment on some general results that we can infer from the definitions and the calculations carried on in the previous paragraphs and sections. In particular, we introduced the generalised page zero \eqref{GCB} that keeps both the modules described above into account. Let us consider $\text{dim} \left( \mathfrak{g} \right) = (a|b)$, $\text{dim} \left( \mathfrak{h} \right) = (r|s)$ (we assume $s \neq 0$) and then $\text{dim} \left( \mathfrak{k} \right) = (a-r|b-s)$. First of all, we remark some facts on the filtrations \eqref{GCA} and \eqref{GCAA} which depend on the picture number $l$, with respect to the odd dimension of the sub-superalgebra. In particular, we have
\begin{equation}\label{GCI}
	F^p \Omega^{(q|l)} \left( \mathfrak{g} \right) = \left\lbrace 0 \right\rbrace \ , \ \forall l > b-s , \forall p,q \in \mathbb{Z} \ .
\end{equation}
This is a consequence of the fact that for $l>b-s$ the space $\Omega^{(q|l)} \left( \mathfrak{g} \right)$ inevitably contains some pseudoforms of $\mathfrak{h}$:
\begin{equation}\label{GCJ}
	\Omega^{(q|l)} \left( \mathfrak{g} \right) = \bigoplus_{i=0}^{b-s} \Omega^{(\bullet|l-i)} \left( \mathfrak{h} \right) \otimes \Omega^{(q-\bullet|i)} \left( \mathfrak{k} \right) \ .
\end{equation}
Hence the filtration is empty, in analogy to \eqref{GCD}. On the other hand, we have
\begin{equation}\label{GCK}
	\tilde{F}^p \Omega^{(q|l)} \left( \mathfrak{g} \right) = \left\lbrace 0 \right\rbrace \ , \ \forall l < s , \forall p,q \in \mathbb{Z} \ ,
\end{equation}
since $\Omega^{(q|l)} \left( \mathfrak{g} \right)$ does not receive any contribution from integral forms of $\mathfrak{h}$, in analogy to \eqref{GCC}. Then, if $\displaystyle s > \frac{b}{2}$, it follows that
\begin{equation}\label{GCL}
	F^p \Omega^{(q|l)} \left( \mathfrak{g} \right) = \left\lbrace 0 \right\rbrace = \tilde{F}^p \Omega^{(q|l)} \left( \mathfrak{g} \right) \ , \ \forall \ b-s < l < s \ .
\end{equation}

In the specific cases described above, the general page zero \eqref{GCB} can be simplified. In the following, we will consider the case where no simplification occurs, namely $\displaystyle s \leq \frac{b}{2}$ and $s \leq l \leq b-s$; the other cases will follow from analogous manipulations. In these cases we have
\begin{equation}\label{GCM}
	F^p \Omega^{(q|l)} \left( \mathfrak{g} \right) = \Omega^{(q|l)}_{q-p} \left( \mathfrak{g} \right) \defeq \bigoplus_{i=0}^{q-p} \Omega^{(q-i|l)} \left( \mathfrak{k} \right) \otimes \Omega^{(i|0)} \left( \mathfrak{h} \right) \ ,
\end{equation}
i.e., using the same notation as in \eqref{TSSH}, the spaces of $(q|l)$-forms which depend \emph{at most} on $q-p$-superforms in $\mathfrak{h}^*$. Analogously,
\begin{equation}\label{GCN}
	\tilde{F}^p \Omega^{(q|l)} \left( \mathfrak{g} \right) = \Omega^{(q|l)}_{\ast 2q-p-r} \left( \mathfrak{g} \right) \defeq \bigoplus_{i=0}^{q-p} \Omega^{(q-r+i|l-s)} \left( \mathfrak{k} \right) \otimes \Omega^{(r-i|s)} \left( \mathfrak{h} \right) \ ,
\end{equation}
where the subscript $2q-p-r$ denotes the highest form number given by pseudoforms in $\mathfrak{k}$, in analogy to \eqref{TSSZG}. \eqref{GCN} can be more conveniently described by noticing that, in analogy to the interpretation of \eqref{TSSZG}, $\Omega^{(r-i|s)} \left( \mathfrak{h} \right)$ is obtained by acting with $i$ contractions on forms of the Berezin bundle of the sub-superalgebra $\mathfrak{h}$ (namely, one has $\Omega^{(r-i|l)} \left( \mathfrak{h} \right) \defeq \mathpzc{B}er \left( \mathfrak{h} \right) \otimes Sym^{i} \Pi \mathfrak{h}$). Thus, $\Omega^{(q|l)}_{\ast 2q-p-r} \left( \mathfrak{g} \right)$ is a space depending \emph{at most} on $q-p$ contractions along vectors in $\mathfrak{h}$.

From the explicit expressions in \eqref{GCM} and \eqref{GCN} we can directly write page 0, which is reported in Table \eqref{TableGCA}:
\begin{equation}\label{GCO}
	\mathcal{E}_0^{m,n} = \left[ \Omega^{(m|l)} \left( \mathfrak{k} \right) \otimes \Omega^{(n|0)} \left( \mathfrak{h} \right) \right] \oplus \left[ \Omega^{(m|l-s)} \left( \mathfrak{k} \right) \otimes \Omega^{(n|s)} \left( \mathfrak{h} \right) \right] \ .
\end{equation}

\begin{table}[ht!]
\centering
\resizebox{1.07\textwidth}{!}{\hspace{-1.4cm}
\begin{tabular}{cccccc}
\multicolumn{1}{c|}{$\ldots$} & \multicolumn{1}{c|}{$\ldots$}                                                                                            & \multicolumn{1}{c|}{$\vdots$} & \multicolumn{1}{c|}{$\ldots$}                                                                                                                                                                            & \multicolumn{1}{c|}{$\ldots$}                                                                                                                                                                                                                               & $\ldots$ \\ \cline{1-2} \cline{4-6} 
\multicolumn{1}{c|}{$\ldots$} & \multicolumn{1}{c|}{$\Omega^{(2|l-s)} \left( \mathfrak{k} \right) \otimes \Omega^{(-1|s)} \left( \mathfrak{h} \right)$}  & \multicolumn{1}{c|}{2}        & \multicolumn{1}{c|}{$\left[ \Omega^{(2|l)} \left( \mathfrak{k} \right) \right] \oplus \left[ \Omega^{(2|l-s)} \left( \mathfrak{k} \right) \otimes \Omega^{(0|s)} \left( \mathfrak{h} \right) \right]$}   & \multicolumn{1}{c|}{$\left[ \Omega^{(2|l)} \left( \mathfrak{k} \right) \otimes \Omega^{(1|0)} \left( \mathfrak{h} \right) \right] \oplus \left[ \Omega^{(2|l-s)} \left( \mathfrak{k} \right) \otimes \Omega^{(1|s)} \left( \mathfrak{h} \right) \right]$}   & $\ldots$ \\ \cline{1-2} \cline{4-6} 
\multicolumn{1}{c|}{$\ldots$} & \multicolumn{1}{c|}{$\Omega^{(1|l-s)} \left( \mathfrak{k} \right) \otimes \Omega^{(-1|s)} \left( \mathfrak{h} \right)$}  & \multicolumn{1}{c|}{1}        & \multicolumn{1}{c|}{$\left[ \Omega^{(1|l)} \left( \mathfrak{k} \right) \right] \oplus \left[ \Omega^{(1|l-s)} \left( \mathfrak{k} \right) \otimes \Omega^{(0|s)} \left( \mathfrak{h} \right) \right]$}   & \multicolumn{1}{c|}{$\left[ \Omega^{(1|l)} \left( \mathfrak{k} \right) \otimes \Omega^{(1|0)} \left( \mathfrak{h} \right) \right] \oplus \left[ \Omega^{(1|l-s)} \left( \mathfrak{k} \right) \otimes \Omega^{(1|s)} \left( \mathfrak{h} \right) \right]$}   & $\ldots$ \\ \cline{1-2} \cline{4-6} 
\multicolumn{1}{c|}{$\ldots$} & \multicolumn{1}{c|}{$\Omega^{(0|l-s)} \left( \mathfrak{k} \right) \otimes \Omega^{(-1|s)} \left( \mathfrak{h} \right)$}  & \multicolumn{1}{c|}{0}        & \multicolumn{1}{c|}{$\left[ \Omega^{(0|l)} \left( \mathfrak{k} \right) \right] \oplus \left[ \Omega^{(0|l-s)} \left( \mathfrak{k} \right) \otimes \Omega^{(0|s)} \left( \mathfrak{h} \right) \right]$}   & \multicolumn{1}{c|}{$\left[ \Omega^{(0|l)} \left( \mathfrak{k} \right) \otimes \Omega^{(1|0)} \left( \mathfrak{h} \right) \right] \oplus \left[ \Omega^{(0|l-s)} \left( \mathfrak{k} \right) \otimes \Omega^{(1|s)} \left( \mathfrak{h} \right) \right]$}   & $\ldots$ \\ \cline{1-2} \cline{4-6} 
$\ldots$                      & -1                                                                                                                       &                               & 0                                                                                                                                                                                                        & 1                                                                                                                                                                                                                                                           & $\ldots$ \\ \cline{1-2} \cline{4-6} 
\multicolumn{1}{c|}{$\ldots$} & \multicolumn{1}{c|}{$\Omega^{(-1|l-s)} \left( \mathfrak{k} \right) \otimes \Omega^{(-1|s)} \left( \mathfrak{h} \right)$} & \multicolumn{1}{c|}{-1}       & \multicolumn{1}{c|}{$\left[ \Omega^{(-1|l)} \left( \mathfrak{k} \right) \right] \oplus \left[ \Omega^{(-1|l-s)} \left( \mathfrak{k} \right) \otimes \Omega^{(0|s)} \left( \mathfrak{h} \right) \right]$} & \multicolumn{1}{c|}{$\left[ \Omega^{(-1|l)} \left( \mathfrak{k} \right) \otimes \Omega^{(1|0)} \left( \mathfrak{h} \right) \right] \oplus \left[ \Omega^{(-1|l-s)} \left( \mathfrak{k} \right) \otimes \Omega^{(1|s)} \left( \mathfrak{h} \right) \right]$} & $\ldots$ \\ \cline{1-2} \cline{4-6} 
\multicolumn{1}{c|}{$\ldots$} & \multicolumn{1}{c|}{$\Omega^{(-2|l-s)} \left( \mathfrak{k} \right) \otimes \Omega^{(-1|s)} \left( \mathfrak{h} \right)$} & \multicolumn{1}{c|}{-1}       & \multicolumn{1}{c|}{$\left[ \Omega^{(-2|l)} \left( \mathfrak{k} \right) \right] \oplus \left[ \Omega^{(-2|l-s)} \left( \mathfrak{k} \right) \otimes \Omega^{(0|s)} \left( \mathfrak{h} \right) \right]$} & \multicolumn{1}{c|}{$\left[ \Omega^{(-2|l)} \left( \mathfrak{k} \right) \otimes \Omega^{(1|0)} \left( \mathfrak{h} \right) \right] \oplus \left[ \Omega^{(-2|l-s)} \left( \mathfrak{k} \right) \otimes \Omega^{(1|s)} \left( \mathfrak{h} \right) \right]$} & $\ldots$ \\ \cline{1-2} \cline{4-6} 
\multicolumn{1}{c|}{$\ldots$} & \multicolumn{1}{c|}{$\ldots$}                                                                                            & \multicolumn{1}{c|}{$\vdots$} & \multicolumn{1}{c|}{$\ldots$}                                                                                                                                                                            & \multicolumn{1}{c|}{$\ldots$}                                                                                                                                                                                                                               & $\ldots$
\end{tabular}
}
\caption{$\mathcal{E}_0^{m,n}$ as defined in \eqref{GCB}. The integers $n$ and $m$ are spanned on the horizontal and vertical axes, respectively.}\label{TableGCA}
\end{table}

The computation of the following pages follows in exact analogy to the example presented in the previous section: the general CE differential reads as in \eqref{TSST}, and in particular the first differential, the horizontal one, formally reads as
\begin{equation}\label{GCP}
	d_0 = V_{\mathfrak{h}} V_{\mathfrak{h}} \iota_{\mathfrak{h}} + V_{\mathfrak{k}} V_{\mathfrak{h}} \iota_{\mathfrak{k}} \ .
\end{equation}
Page 1 of the spectral sequence is naturally defined, for any picture number $l$, as
\begin{equation}\label{GCQ}
	\mathcal{E}_1^{m,n} \defeq H \left( \mathcal{E}_0^{m,n} , d_0 \right) \ .
\end{equation}
By following the same arguments of the previous section, we directly see that the part $V_{\mathfrak{h}} V_{\mathfrak{h}} \iota_{\mathfrak{h}}$ of $d_0$ selects the cohomology spaces $\displaystyle H^{(n|\bullet)} \left( \mathfrak{h} \right) , \bullet = 0 , s $. On the other hand, the part $V_{\mathfrak{k}} V_{\mathfrak{h}} \iota_{\mathfrak{k}}$ selects the invariant forms among the forms in $\mathfrak{k}$, i.e., $\displaystyle \left( \Omega^{(m|\bullet)} \left( \mathfrak{k} \right) \right)^{\mathfrak{h}}$. Then we have
\begin{equation}\label{GCR}
	\mathcal{E}_1^{m,n} = \left[ \left( \Omega^{(m|l)} \left( \mathfrak{k} \right) \right)^{\mathfrak{h}} \otimes H^{(n|0)} \left( \mathfrak{h} \right) \right] \oplus \left[ \left( \Omega^{(m|l-s)} \left( \mathfrak{k} \right) \right)^{\mathfrak{h}} \otimes H^{(n|s)} \left( \mathfrak{h} \right) \right] \ .
\end{equation}
Notice that both spaces of \eqref{GCR} are zero for $n < 0$: $H^{(n|0)} \left( \mathfrak{h} \right) = \left\lbrace 0 \right\rbrace , \forall n \in \mathbb{Z}, n < 0$ follows trivially from the fact that superforms are bounded from below; $H^{(n|s)} \left( \mathfrak{h} \right) = \left\lbrace 0 \right\rbrace , \forall n \in \mathbb{Z}, n < 0$ follows from the fact that $H^{(n|0)} \left( \mathfrak{g} \right) \neq \left\lbrace 0 \right\rbrace \iff n \leq \text{dim}_0 \left( \mathfrak{g} \right)$ for any basic Lie superalgebra $\mathfrak{g}$, then, from the Berezinian complement isomorphism \eqref{VSPQ} it follows that $H^{(n|s)} \left( \mathfrak{g} \right) \neq \left\lbrace 0 \right\rbrace \iff n \geq 0$. This means that in the tabular representation of page 1, the second and third quadrants are empty.

The following step is the definition of page 2: the differential $d_1$, i.e., the vertical differential, in general formally reads as
\begin{equation}\label{GCS}
	d_1 = V_{\mathfrak{k}} V_{\mathfrak{k}} \iota_{\mathfrak{k}} + V_{\mathfrak{k}} V_{\mathfrak{h}} \iota_{\mathfrak{h}} \ ,
\end{equation}
and page 2 is defined as
\begin{equation}\label{GCT}
	\mathcal{E}_2^{m,n} \defeq H \left( \mathcal{E}_1^{m,n} , d_1 \right) \ .
\end{equation}
The computation of page 2 is an hard task in general, but it simplifies greatly for reductive Lie sub-superalgebras, since the second differential becomes $d_1 = V_{\mathfrak{k}} V_{\mathfrak{k}} \iota_{\mathfrak{k}}$, corresponding to the relative differential for the coset $\mathfrak{k}$. For reductive Lie sub-superalgebras we have that $d_1$ acts on $\displaystyle \left( \Omega^{(m|l)} \left( \mathfrak{k} \right) \right)^{\mathfrak{h}} \otimes H^{(n|0)} \left( \mathfrak{h} \right)$ only, giving the relative cohomologies. Hence, page 2 reads as
\begin{equation}\label{GCU}
	\mathcal{E}_2^{m,n} = \left[ H^{(m|l)} \left( \mathfrak{g} , \mathfrak{h} \right) \otimes H^{(n|0)} \left( \mathfrak{h} \right) \right] \oplus \left[ H^{(m|l-s)} \left( \mathfrak{g} , \mathfrak{h} \right) \otimes H^{(n|s)} \left( \mathfrak{h} \right) \right] \ .
\end{equation}
We want to emphasise that the results obtained in \eqref{GCO}, \eqref{GCR} and \eqref{GCU} mirror directly the results reported in theorems 15.1, 15.2 and 15.3 of \cite{Koszul}, respectively. Really, for reductive Lie sub-superalgebras, we have that \eqref{GCU} is the super Lie algebra extension of page 2 as calculated for Lie algebras (w.r.t. reductive Lie sub-algebras), constructed with the relative cohomology spaces and the cohomology spaces of the sub-algebra. The key point is, once again, that the algebraic cohomology of a Lie superalgebra is distributed at various picture numbers, in general among pseudoforms as well as among superforms and integral forms.

\section{Cohomology via Explicit Realisation}

After the formal discussion on the cohomology, we present here a complete and explicit realisation of it for the example of $\mathfrak{osp}(2|2)$. We first discuss the result using the associated Poincar\'e polynomials and then 
we provide an explicit construction of the class representatives. In appendix A we collect some basic computation rules concerning the distributional realisation of integral forms and pseudoforms.

\subsection{Poincar\'e Polynomials}

We briefly review the definition of Poincar\'e series and Poincar\'e polynomials. 
For $X$ a \emph{graded} $\mathbb{K}$-vector space with direct decomposition into $p$-degree homogeneous subspaces given by 
$X = \bigoplus_{p \in \mathbb{Z}} X_p$ we call the formal series 
\begin{eqnarray}
\label{POIA}
\mathpzc{P}_X(t) = \sum_p ({\rm dim}_k \, X_p) (-t)^p
\end{eqnarray}
the \emph{Poincar\'e series} of $X$. Notice that we have implicitly assumed that $X$ is of \emph{finite type}, \emph{i.e.} its homogeneous subspaces $X_p$ are finite dimensional for every $p.$ The unconventional sign in $(-t)^p$ takes into account the \emph{parity} of $X_p$, which takes values in $\mathbb{Z}_2$ and it is given by $p \, \mbox{mod}\, 2$; this is particularly useful for super algebras. %If also $\dim_\mathbb{K} X$ is finite, then $\mathpzc{P}_X(t)$ becomes a polynomial $\mathpzc{P}_X [t]$, called \emph{Poincar\'e polynomial} of $X$. 

If we assume that the pair $(X, \delta)$ is a differential complex for a graded vector space $X$ and $\delta : X_p \rightarrow X_{p+1}$ for any $p$, then the cohomology $H_{\delta}^\bullet (X) = \bigoplus_{p\in \mathbb{Z}} H_{\delta}^p(X)$ is a graded space. The numbers $b_p (M) \defeq \dim_\mathbb{K} H^p_{dR} (M)$ are identified with the Betti numbers of 
a given manifold and, here, by extension, we will call \emph{Betti numbers} the dimensions of any cohomology space valued in a field. In particular, we will call $p$-th Betti numbers of a certain Lie (super)algebra the dimension of its Chevalley-Eilenberg $p$-cohomology group $b_p (\mathfrak{g}) = \dim_\mathbb{K} H^p_{CE} (\mathfrak{g}),$ so that the Poincar\'e series of the Lie (super)algebra $\mathfrak{g}$ is the generating function of its Betti numbers:
\bear
\mathpzc{P}_{\mathfrak{g}} (t) = \sum_p b_p (\mathfrak{g}) (-t)^p.
\eear
Notice that we used the word \virgolette series" on purpose: indeed, $H^\bullet_{CE} (\mathfrak{g})$ is not in general finite dimensional for a generic Lie superalgebra $\mathfrak{g}$ (see, e.g., \cite{CCGN2}). Sometimes, it is useful to introduce a second grading. In that case the space is said to be 
{\it bigraded vector space} $X = \sum_{p,q \in \mathbb{Z}} \Omega^{(p|q)}$ where the two numbers $(p|q)$ denote the form number and the 
picture number, then we can write a {\it double Poincar\'e series} 
\begin{eqnarray}
\label{POIE}
\mathpzc{P}_X(t,\tilde t) = \sum_{p,q} (-t)^p (-\tilde t)^q {\rm dim}\Omega^{(p|q)} \ ,
\end{eqnarray}
 which, in any case, allows an easier identification of cohomology spaces.

First, let us recollect some results from \cite{CE}. If one considers the bosonic sub-algebra $\mathfrak{so}(2) \oplus \mathfrak{sp}(2)$ of 
$\mathfrak{osp}(2|2)$, its Poincar\'e polynomial is factorized into a product of two polynomials (as follows from the K\"unneth formula) 
\begin{eqnarray}
\label{PPA}
\mathpzc{P}_{\mathfrak{so}(2) \oplus \mathfrak{sp}(2)}(t) = (1- t) (1-t^3) \ ,
\end{eqnarray}
counting both the cohomology classes of the abelian factor $\mathfrak{so}(2)$ and those of the non-abelian one $\mathfrak{sp}(2)$. In the same 
way, for the super sub-algebra $\mathfrak{osp}(1|2)$ we have 
\begin{eqnarray}
\label{PPB}
\mathpzc{P}_{\mathfrak{osp}(1|2)}(t,\tilde t) = (1-t^3) (1 + \tilde t^2)   = (1-t^3) +  (1-t^3) \tilde t^2\,,  
\end{eqnarray}
where the polynomial $(1-t^3)$ counts the cohomology spaces of superforms $H^\bullet_{super}(\mathfrak{osp}(1|2))$ while 
$(1-t^3) \tilde t^2$ takes into account the cohomology spaces of integral forms $H^\bullet_{integral}(\mathfrak{osp}(1|2))$. The term $\tilde t^2$ 
takes into account the picture of the integral forms, here the maximal picture is two. The Berezinian complement duality constructed in \cite{CCGN2} relates $H^\bullet_{super}(\mathfrak{osp}(1|2))$ 
to $H^\bullet_{integral}(\mathfrak{osp}(1|2))$ which are isomorphic.

Poincar\'e series turn out to be a particularly useful tool when dealing with cosets: in \cite{GHV} there is the proof of a theorem that allows to calculate the Poincar\'e polynomial of certain cosets, without explicitly calculating all the cohomology representatives. Given a Lie algebra $\mathfrak{g}$ with Poincar\'e polynomial $\mathpzc{P}_\mathfrak{g} (t) = \sum_i \left( 1 - t^{c^{\mathfrak{g}}_i} \right)$, where $c^{\mathfrak{g}}_i$ are the usual exponents in the factorised form of the polynomial, and $\mathfrak{h}$ a Lie sub-algebra of $\mathfrak{g}$, of the same rank, with Poincar\'e polynomial given by $\mathpzc{P}_\mathfrak{h} (t) = \sum_i \left( 1 - t^{c^{\mathfrak{h}}_i} \right)$, then the Poincar\'e polynomial for the coset will be given by
\bear \label{cosetp}
\mathpzc{P}_{\mathfrak{g}/\mathfrak{h}}(t) = \frac{\prod_{i} (1-t^{c^\mathfrak{g}_i+1})}{\prod_{j} (1-t^{c^\mathfrak{h}_j+1})} \ .
\eear
Actually, since we are dealing with superalgebras, we have to adapt the previous notions in order to keep into account for the picture number. This means that the Poincar\'e polynomial of $\mathfrak{g}$ and $\mathfrak{h}$ will read in general $\mathpzc{P}_\mathfrak{g} (t, \tilde{t}) = \sum_i \left( 1 - t^{c^{\mathfrak{g}}_i} \tilde{t}^{p^{\mathfrak{g}}_i} \right)$ and $\mathpzc{P}_\mathfrak{h} (t, \tilde{t}) = \sum_i \left( 1 - t^{c^{\mathfrak{h}}_i} \tilde{t}^{p^{\mathfrak{h}}_i} \right)$, respectively. We can then extend \eqref{cosetp} to superalgebras as
\begin{equation}\label{cosetpA}
	\mathpzc{P}_{\mathfrak{g}/\mathfrak{h}}(t, \tilde{t}) = \frac{\prod_{i} \left( 1-t^{c^\mathfrak{g}_i+1} \tilde{t}^{p^{\mathfrak{g}}_i} \right)}{\prod_{j} \left(1-t^{c^\mathfrak{h}_j+1} \tilde{t}^{p^{\mathfrak{g}}_j} \right)} \ .
\end{equation}
This product formula is very helpful since it provides some informations regarding the 
different cohomology classes, as we will show for our main example in the following. In App. B we also report the application of \eqref{cosetpA} to two cosets of $\mathfrak{osp}(1|2)$.

\subsection{Explicit Construction}

Let us now move to the main example. Collecting the results of the second section into the Poincar\'e polynomial of $\mathfrak{osp}(2|2)$, we get 
\begin{eqnarray}
\label{PPD}
\mathpzc{P}_{\mathfrak{osp}(2|2)}(t,\tilde t) &=& (1- t^3) (1 - t \tilde t^2) (1 + \tilde t^2) = \nonumber \\&=& (1- t^3) - (1 - t^3) t \tilde t^2 + (1-t^3) \tilde t^2  - (1-t^3) t \tilde t^4 \ .
\end{eqnarray}
Reading the polynomial, the first parenthesis represents superform cohomology $H^{(p|0)}$ with $p=0$ and $p=3$ (see \cite{Fuks}). The second polynomial 
$- t \tilde t^2 + t^4 \tilde t^2$ corresponds to the pseudoform cohomology $H^{(p|2)}$, with $p=1$ and $p=4$, discussed in \eqref{TSSY}
and selected with the filtration \eqref{TSSA}. The third polynomial $\tilde t^2 -t^3 \tilde t^2$ corresponds to the pseudoform cohomology selected by the inequivalent 
filtration \eqref{TSSZA} and discussed in \eqref{TSSZM}. Finally, the last polynomial $-  t \tilde t^4 + t^4\tilde t^4$ counts the cohomology of 
integral forms $H^{(p|4)} \left( \mathfrak{g} \right)$. The latter is isomorphic to $H^{(p|0)}$, because of the Berezinian complement duality $\star H^{(p|0)} = H^{(4-p|4)}$, as discussed in the previous sections. In addition, we noticed that 
this duality holds among pseudoforms as well: $\star H^{(p|2)} \left( \mathfrak{g} \right) = H^{(4-p|2)} \left( \mathfrak{g} \right)$. %The Poincar\'e polynomial \eqref{PPD} takes cares of the parity of the different classes. 

Let us fix $\mathfrak{h} = \mathfrak{g}_0 = \mathfrak{so}(2) \times \mathfrak{sp}(2)$. We now study the coset space $\mathfrak{k} = \mathfrak{g/h}$ 
in order to describe the cohomology of $\mathfrak{g}$ by using the explicit distributional realisation of pseudoforms and integral forms. For that, we compute the Poincar\'e polynomial of $\mathfrak{k}$ using \eqref{cosetpA}:
\begin{eqnarray}
\label{PPE}
\mathpzc{P}_{\mathfrak{k}}(t,\tilde t) 
&=& \frac{(1- t^4) (1 - t^2 \tilde t^2) (1 + \tilde t^2)}{(1-t^2) (1-t^4)} 
=  \frac{(1 - t^2 \tilde t^2) (1 + \tilde t^2)}{(1-t^2)} =
\nonumber \\
&=& 
 \frac{1 + (1-t^2) t^2 \tilde t^2  - t^2 \tilde t^4}{(1-t^2)} = 
\frac{1}{1-t^2} + \tilde t^2 + \frac{1}{1- \frac{1}{t^2}} \tilde t^4 \ .
\end{eqnarray}
 The coset space $\mathfrak{k}$ is a $(0|4)$-dimensional space, with no bosonic 
 coordinates. Therefore 
 the MC forms $\psi^\pm, \bar\psi^\pm$ are covariantly constant, with respect the covariant differential $\nabla$, so that the 
 equivariant cohomology $H_{CE}(\mathfrak{k})$ is easily computed. The 
 three pieces of the series \eqref{PPE} corresponds to $H_{super}^\bullet(\mathfrak{k}), 
 H_{psuedo}^\bullet(\mathfrak{k})$ and $H_{integral}^\bullet(\mathfrak{k})$ and here 
 we discuss them explicitly. 
 
 The cohomology of superforms $H_{super}^\bullet(\mathfrak{k})$ is generated  by any power of the $(2|0)$-form (which is basic, according to \eqref{VSPN})
\begin{eqnarray}
\label{PPF}
K^{(2|0)} = \psi^+\wedge \bar{\psi}^- - \bar{\psi}^+ \wedge \psi^- \ , ~~~~
\nabla K^{(2|0)} =0\,.
\end{eqnarray}
Note that, according to \eqref{VSPJ}, $K^{(2|0)} \propto dU$, so that $\mathfrak{h}$-invariance is straightforwardly verified. Since $K^{(2|0)}$ is an even cohomology representative, it is easy to see that any power of it is taken into account in the Poincar\'e 
series as $\sum_{p\leq 0} t^{2p} = 1/(1- t^2)$, corresponding to the first term of \eqref{PPE}.

$K_2$ can also be written as $K_2 = \epsilon^{\a\b} \epsilon_{IJ} \psi^I_\alpha \psi^J_\beta$, where $\psi^I_\alpha$ are the MC forms in the real representation ($I,J=1,2$ and 
$\alpha, \beta =1,2$):
\begin{equation}
	\psi^1_1 = \psi^+ \ , \ \psi^1_2 = \psi^- \ , \ \psi^2_1 = \bar{\psi}^+ \ , \ \psi^2_2 = \bar{\psi}^- \ .
\end{equation} 

The cohomology of integral forms is computed by using the Berezinian duality prescription as follows: we start from the 
top integral form of the supercoset $\mathfrak{k}$, which explicitly reads
\begin{eqnarray}
\label{PPG}
\omega^{(0|4)} = \delta(\psi^+) \delta(\psi^+) \delta(\bar\psi^+) \delta(\bar\psi^+) \ .
\end{eqnarray}
\eqref{PPG} is covariantly closed $\nabla \omega^{(0|4)}=0$ and basic. Analogously, any other integral form obtained by acting with derivatives 
on Dirac delta functions is also covariantly closed. Then, we should select the basic ones. By respecting the symmetry of the sub-algebra, we define the contraction 
operator $\iota_2$ as
\begin{eqnarray}
\label{PPH}
\iota_2 = \iota_{F^+} \iota_{\bar{F}^-} - \iota_{\bar F^+} \iota_{F^-} \ ,
\end{eqnarray}
where $F^\pm$ and $\bar F^\pm$ are the odd generators of the 
superalgebra. $\iota_2$ is defined to be the formal inverse (modulo multiplication by constants) of $K^{(2|0)}$. We can act with any power $\iota_2^p$ on $\omega^{(0|4)}$ 
to get the infinite number of cohomology representatives 
\begin{eqnarray}
\label{PPI}
\iota^{p}_2 \delta(\psi^+) \delta(\psi^+) \delta(\bar\psi^+) \delta(\bar\psi^+) \ , \forall p \geq 0 \ ,
\end{eqnarray}
which are covariantly constant and generate the complete integral form cohomology
$H_{integral}^\bullet \left( \mathfrak{k} \right)$. Indeed, since $\iota_2^p$ corresponds to 
$t^{-2p}$, the Poincar\'e series for integral forms reads 
\begin{eqnarray}
\label{PPJ}
\sum_{p=0}^\infty t^{-2p} \tilde{t}^4 = \frac{1}{(1 - \frac{1}{t^2})} \tilde t^4 \ . %= - \frac{1}{1-t^2} t^2 \tilde t^4 \ .
\end{eqnarray}

Let us come to the final piece of \eqref{PPE}, namely the single term $\tilde t^2$. 
This means that there is a single cohomology generator among pseudoforms. 
In order to single it out, we have to construct 
a pseudoform which is $\mathfrak{so}(2) \oplus \mathfrak{sp}(2)$ invariant. 

By using the real representation $\psi^I_\alpha$, we observe that the combinations 
\begin{eqnarray}
\label{PPM}
\eta(\psi^1) = \epsilon_{\a\b}\delta(\psi^1_\alpha) \delta(\psi^1_\beta)\,, ~~~~
\eta(\psi^2) = \epsilon_{\a\b}\delta(\psi^2_\alpha) \delta(\psi^2_\beta)\,, ~~~~
\end{eqnarray}
are invariant w.r.t. the sub-algebra $\mathfrak{sp}(2)$: given $X \in \mathfrak{sp}(2)$, we have
\begin{eqnarray}
	\mathcal{L}_X \eta(\psi^1) = \left( \iota_X d + d \iota_X \right) 2 \delta(\psi^1_1) \delta(\psi^1_2) = \iota_X \left[ U \psi^2_1 \iota^1_1 \delta(\psi^1_1) \delta(\psi^1_2) + U \psi^2_2 \iota^1_2 \delta(\psi^1_1) \delta(\psi^1_2) \right] = 0 \ ,
\end{eqnarray}
where we have used \eqref{VSPG} $\div$ \eqref{VSPK}. The same holds true for $\eta(\psi^2)$.

The invariance under $\mathfrak{so}(2)$ is implemented by 
requiring
\begin{eqnarray}
\label{PPN}
\sum_{\alpha=1,2} \Big( 
\psi^1_\alpha \iota_{F^2_\alpha} - \psi^2_\alpha \iota_{F^1_\alpha} 
\Big) \sigma(\psi^1, \psi^2) =0 \ ,
\end{eqnarray}
where $F^I_\alpha$ are the dual vectors to $\psi^I_\alpha$: $ \iota_{F^I_\alpha} \psi^J_\beta = \delta^J_I \delta^\alpha_\beta$. The generalized form $\sigma(\psi^1, \psi^2)$ satisfying \eqref{PPN}
depends on $\psi^I_\alpha$ through the combinations in \eqref{PPM}. 
Note that,  being $F^I_\alpha$ odd, the differential operators 
$\iota_{F^I_\alpha}$ are even.
The solution of \eqref{PPN} can be expressed in terms of the formal
$0^{th}$-order Bessel function $J_0(x) = 1 - x^2/4 + x^4/64 + \dots $ as follows:
\begin{eqnarray}
\label{PPP}
\sigma(\psi^1, \psi^2) = J_0
\Big( \sum_{\alpha} \psi^2_\a \iota_{F^1_\alpha} \Big) \eta(\psi^1) = 
\eta(\psi^1) - \frac14  \sum_{\alpha,\beta} \psi^2_\a  \psi^2_\beta \iota_{F^1_\alpha} 
\iota_{F^1_\beta}   \eta(\psi^1) + \dots \ .
\end{eqnarray}
The invariant pseudoform $\sigma(\psi^1, \psi^2)$ can also be expanded around 
$\eta(\psi^2)$, with an analogous expression that exchanges $\psi^1$ with $\psi^2$:
\begin{equation}\label{PPPA}
	\sigma(\psi^1, \psi^2) = J_0
\Big( \sum_{\alpha} \psi^1_\a \iota_{F^2_\alpha} \Big) \eta(\psi^2) = 
\eta(\psi^2) - \frac14  \sum_{\alpha,\beta} \psi^1_\a  \psi^1_\beta \iota_{F^2_\alpha} 
\iota_{F^2_\beta} \eta(\psi^2) + \dots \ .
\end{equation} 
The equivalence of the two representations is consistent with the Berezinian complement duality, since the unique pseudoform counted by the monomial $\tilde t^2$ in the Poincar\'e polynomial \eqref{PPD} is self-dual, as explicitly shown with the representatives \eqref{PPP} and \eqref{PPPA}. 

The results of this section are highly non-trivial: we have shown that the explicit distributional realisation of pseudoforms and integral forms is a powerful tool to calculate cohomology representatives and that the explicit results are consistent with those obtained without referring to any realisation through spectral sequences. Nonetheless, we have been able to construct a pseudoform which is invariant w.r.t. $\mathfrak{so}(2) \oplus \mathfrak{sp}(2)$; the explicit realisation, supported by the abstract counterpart, could serve as a starting point for the introduction of pseudoforms in more general contexts, e.g., supermanifolds: one could think to introduce pseudoforms as integral forms of sub-supermanifolds respecting some isometries. The general setup for these definitions is once again suggested by the algebraic setting analysed in this paper: relative (co)homology. This will be subject of future investigations.

\section*{Acknowledgements}
\noindent This work has been partially supported by Universit\`a del Piemonte Orientale research funds. We thank S.Cacciatori, R. Catenacci and S. Noja for many useful discussions.

\appendix
\section{A Brief Review on Integral Forms}

In this appendix we recall the main computation techniques for integral forms in the distributional realisation. For a more exhaustive review we suggest \cite{CCGN2} for their introduction in the superalgebraic setting and \cite{CG,CG2, Witten} for their use on supermanifolds.

We consider a supermanifold ${\cal SM}^{(m|n)}$ with $m$ bosonic and $n$ fermionic dimensions. We denote the local coordinates in an open set as $(x^a, \theta^\alpha), a=1,\ldots,m , \alpha=1,\ldots,n$. A generic integral form locally reads 
\begin{equation}\label{ABRIFD}
	\omega^{(p|n)} = \omega_{[i_1 \ldots i_r]}^{(\alpha_1 \ldots \alpha_s)} \left( x , \theta \right) dx^{i_1} \wedge \ldots \wedge dx^{i_r} \wedge \iota_{\alpha_1} \ldots \iota_{\alpha_s} \delta \left( d \theta^1 \right) \wedge \ldots \wedge \delta \left( d \theta^n \right) \ ,
\end{equation}
where $\delta \left( d \theta \right)$ are (formal) Dirac delta distributions and $\iota_\alpha$ denotes the interior product. The symbol  $\delta \left( d \theta \right)$ satisfies the following distributional identities
\begin{equation}\label{ABRIFEA}
	 d \theta \delta \left( d \theta \right) = 0 \ \ , \ \ \delta \left( \lambda d \theta \right) = \frac{1}{\lambda} \delta \left( d \theta \right) \ \ , \ \ d \theta \iota^{(p)} \delta \left( d \theta \right) = - p \delta^{(p-1)} \left( d \theta \right) \ \ , $$ $$
	 \delta \Big( d \theta^\alpha \Big) \wedge \delta \left( d \theta^\beta \right) = - \delta \left( d \theta^\beta \right) \wedge \delta \Big( d \theta^\alpha \Big) \ \ , \ \ dx \wedge \delta \left( d \theta \right) = - \delta \left( d \theta \right) \wedge d x \ \ ,
	 \end{equation}
indicating that actually these are not conventional distributions, but rather \emph{de Rham currents} (see \cite{Witten}). 

Given these properties, we retrieve a top form among integral forms as
\begin{equation}\label{ABRIFF}
	\omega_{top}^{(m|n)} = \omega \left( x , \theta \right) dx^1 \wedge \ldots \wedge dx^m \wedge \delta \left( d \theta^1 \right) \wedge \ldots \wedge \delta \left( d \theta^n \right) \ ,
\end{equation}
where $\omega \left( x , \theta \right)$ is a superfield. The space of $(m|n)$ forms corresponds to 
the \emph{Berezinian bundle}, since  the generator $dx^1 \wedge \ldots \wedge dx^m \wedge \delta \left( d \theta^1 \right) \wedge \ldots \wedge \delta \left( d \theta^n \right)$ transforms under change of coordinates as the superdeterminant of the Jacobian.

One can also consider the classes of forms with non-maximal and non-zero number of delta's: the \emph{pseudoforms}. A general pseudoform with $q$ Dirac delta's (i.e., picture number $q$) is locally given by
\begin{equation}\label{ABRIFG}
	\omega^{(p|q)} = \omega_{[a_1 \ldots a_r](\alpha_1 \ldots \alpha_s)[\beta_1 \ldots \beta_q]} \left( x , \theta \right) dx^{a_1} \wedge \ldots \wedge dx^{a_r} \wedge d \theta^{\alpha_1} \wedge \ldots \wedge d \theta^{\alpha_s} \wedge \delta^{(t_1)} \left( d \theta^{\beta_1} \right) \wedge \ldots \wedge \delta^{(t_q)} \left( d \theta^{\beta_q} \right) \ ,
\end{equation}
where $\delta^{(i)} \left( d \theta \right) \equiv \left( \iota \right)^i \delta \left( d \theta \right)$. The form number of \eqref{ABRIFG} is obtained as 
\begin{equation}\label{ABRIFH}
	p = r + s - \sum_{i=1}^q t_i \ ,
\end{equation}
since the contractions carry negative form number. The two numbers $p$ and $q$ in eq. \eqref{ABRIFH}, corresponding to the form number and the picture number, respectively, range as $-\infty < p < +\infty$ and $0 \leq q \leq n$. If $q=0$, we have superforms, if $q=n$ we have integral forms, if $0<q<n$ we have pseudoforms.

\section{Poincar\'e Polynomials \& Cosets: $\mathfrak{osp}(1|2)$}

In this appendix we want to show an example of the application of Poincar\'e polynomials as a support to calculate the algebraic cohomology of cosets. In particular, we will use \eqref{cosetpA} to compute the cohomology of the two cosets 
$\mathfrak{k}_1 = \mathfrak{osp}(1|2)/\mathfrak{so}(2)$ and  $\mathfrak{k}_2 = \mathfrak{osp}(1|2)/\mathfrak{sp}(2)$. The former is $(2|2)$-dimensional 
and its Poincar\'e polynomial is 
\begin{eqnarray}
\label{PPC}
\mathpzc{P}_{\mathfrak{k}_1}(t,\tilde t)  = \frac{ (1-t^4) (1 + \tilde t^2) }{(1-t^2)} = (1 + t^2) (1 + \tilde t^2) \ ,
\end{eqnarray}
corresponding to the equivariant cohomology of the coset space. The polynomial counts four classes, 
whose representatives are (we use the same notations as in section 1)
\begin{eqnarray}
\label{PPDapp}
\omega^{(0|0)} &=&1 \ , \nonumber \\ 
\omega^{(2|0)} &=& V^{++}\wedge V^{--} + \psi^+\wedge \psi^- \ , \nonumber \\
\omega^{(0|2)} &=& \delta(\psi^+) \delta(\psi^-) + V^{++} \wedge V^{--} \delta'(\psi^+) \delta'(\psi^-) \ ,  \nonumber \\
\omega^{(2|2)} &=& V^{++}\wedge V^{--}  \delta(\psi^+) \delta(\psi^-) \ , 
\end{eqnarray}
which are closed under the $\mathfrak{so}(2)$-covariant CE differential. 

The coset $\mathfrak{k}_2 = \mathfrak{osp}(1|2)/\mathfrak{sp}(2)$ is $(0|2)$-dimensional 
with anticommuting coordinates only. Using again \eqref{cosetpA}, we 
obtain
\begin{eqnarray}
\label{PPEapp}
\mathpzc{P}_{\mathfrak{k}_2}(t,\tilde t)  = \frac{ (1-t^4) (1 + \tilde t^2) }{(1-t^4)} =(1 + \tilde t^2) \ , 
\end{eqnarray}
which is the $\mathfrak{sp}(2)$ equivariant cohomology. As discussed in \cite{CE}, $H^\bullet_{super}(\mathfrak{k}_2)$ is generated by $\omega^{(0|0)}=1$ and 
$H^\bullet_{integral}(\mathfrak{k}_2)$ is generated by the integral form $\omega^{(0|2)}= \delta(\psi^+) \delta(\psi^-)$. The two spaces are isomorphic 
by Berezinian complement duality.

\end{document}